\documentclass[11pt]{article}

\usepackage[margin=1in]{geometry}
\usepackage{amsmath, amssymb, amsthm}
\usepackage{graphicx}
\usepackage{booktabs}
\usepackage{hyperref}
\usepackage[authoryear,round]{natbib}
\bibpunct{(}{)}{;}{a}{}{,}
\usepackage{microtype}
\usepackage{algorithm}
\usepackage{enumitem}
\usepackage{subcaption}
\usepackage{xcolor}
\usepackage{times}
\usepackage{epstopdf}
\usepackage{nomencl}
\usepackage{multicol}
\makenomenclature

\renewcommand{\nompreamble}{\begin{multicols}{2}}
	\renewcommand{\nompostamble}{\end{multicols}}

\DeclareMathOperator*{\argmin}{arg\,min}

\usepackage{pgffor}
\usepackage{setspace}

\makeatletter
\def\fs@ruled{\def\@fs@cfont{\bfseries}\let\@fs@capt\floatc@ruled
	\def\@fs@pre{}\def\@fs@post{\kern2pt\hrule\relax}%
	\def\@fs@mid{\kern2pt\hrule\kern2pt}%
	\let\@fs@iftopcapt\iftrue}
\makeatother

\title{Exact Bayesian Planning for Simple Step-Stress Accelerated Life Testing with Competing Risks}
\author{
	Kiran Prajapat\thanks{Corresponding author. Email: kiranprajapat92@gmail.com, kiran.prajapat@newcastle.ac.uk}\\
	\small School of Mathematics, Statistics and Physics, Newcastle University, NE1 7RU, Newcastle, UK
}\date{}

\begin{document}
	
	\maketitle
	
	\begin{abstract}
		We propose a Bayesian framework for planning simple 
		step-stress accelerated life tests when items are subject 
		to two independent competing failure modes We assume that the competing risks are independent, with 
		lifetimes following Weibull distributions, and adopt the 
		cumulative exposure model with a log-linear stress-life 
		relationship to connect failure time distributions across 
		stress levels. The optimality criterion is the preposterior variance of the $p$-th quantile of the 
		lifetime distribution at use stress, evaluated without reliance on 
		asymptotic approximations, making the methodology valid 
		regardless of sample size. Building on the idea of quantile-based reparametrisation 
		used in single-mode ALT \citep{zhang2006bayesian}, we 
		extend this approach to the competing risks setting by 
		reparametrising the model parameters for each failure mode 
		to physically interpretable and approximately independent 
		quantities, making it possible to elicit priors directly 
		from engineering knowledge of device behaviour. Posterior inference is carried out 
		using the No-U-Turn Sampler implemented in Stan, and 
		the optimal design is located via Monte Carlo simulation 
		over a grid of candidate designs. The methodology is 
		illustrated on a real step-stress dataset for a solar 
		lighting device subject to capacitor and controller 
		failure modes. A comprehensive sensitivity analysis with 
		respect to the quantile probability, the lower stress 
		level, the prior hyperparameter specification, and the 
		sample size shows that the optimal stress-change time 
		is moderately sensitive to these inputs while the 
		optimal lower stress level consistently favours 
		operation close to use conditions, a finding that holds 
		across all prior specifications considered.
	\end{abstract}
	
	[2pt]
	
	%	\newpage
	%	\tableofcontents
	\newpage
	
	\newtheorem{theorem}{Theorem}[section]
	\newtheorem{corollary}{Corollary}[theorem]
	\newtheorem{proposition}{Proposition}[section]
	\newtheorem{lemma}{Lemma}[section]
	\newtheorem{remark}{Remark}[section]
	%	\section{Notations and Abbreviations}
	
	\nomenclature[N]{$k$}{number of risks}%
	\nomenclature[N]{$h$}{number of stress levels}%
	\nomenclature[N]{$n$}{sample size}
	\nomenclature[A]{PDF}{probability density function}
	\nomenclature[A]{CDF}{cumulative distribution function}
	\nomenclature[N]{$s_1, ~ s_2$}{stress levels}
	\nomenclature[N]{$x_1,~ x_2$}{standardized stress levels}
	\nomenclature[N]{$\tau$}{stress changing point}
	\nomenclature[N]{$t_c$}{experimental duration}
	\nomenclature[N]{$n_c$}{total number of failures at the end of the experiment}
	\nomenclature[N]{$T_{i,j}(x)$}{lifetime of an $i$-th unit that failed because of risk-$j$ at stress $x$}%
	\nomenclature[N]{$T_{i}(x)$}{observed lifetime of an $i$-th unit at stress $x$}
	\nomenclature[N]{$a_j, ~ b_j $}{intercept and slope parameters in log-linear form of the stress-lifetime model with competing risk $j$}
	\nomenclature[N]{$\theta$}{scale parameter in the Weibull distribution}
	\nomenclature[N]{$\beta$}{shape parameter in the Weibull distribution}
	\nomenclature[N]{$f_W(.;\theta, \beta)$ }{Weibull PDF with scale $\theta$ and shape $\beta$}
	\nomenclature[N]{$F_W(.;\theta, \beta)$ }{Weibull CDF with scale $\theta$ and shape $\beta$}
	\nomenclature[N]{$f_X(.)$ }{PDF of a random variable $X$}
	\nomenclature[N]{$C$}{a random variable that denotes type of failure}
	\nomenclature[N]{$I(t)$}{index of the stress level $x_{I(t)}$ at time-point $t$}
	\nomenclature[N]{$F_{l,j}$}{sub-distribution function at time-point $t$ for the unit failed due to risk $j$ stress level $x_{l}$}
	\nomenclature[A]{ALT}{accelerated life testing }
	\nomenclature[A]{CSALT}{constant-stress accelerated life testing} 
	\nomenclature[A]{SSALT}{step-stress accelerated life testing}
	\nomenclature[N]{$L$}{likelihood function}
	\nomenclature[N]{$\ell$}{log-likelihood function}
	\nomenclature[N]{$\underline{\boldsymbol{\theta}}$}{vector of the SSALT model-parameters with $\underline{\boldsymbol{\theta}}=(a_1,b_1,\beta_1,a_2,b_2,\beta_2)$ }
	%\nomenclature[N]{$n_{l,j}$}{number of failures because of risk $j$ at stress level $x_l$}
	\nomenclature[N]{$\psi_{l,j}(t)$}{cumulative exposure at stress level $x_l$ for time-point $t$}
	\nomenclature[N]{$\tilde{t}_{l,j}(t)$}{transformed time at $t$ in SSALT at stress level $x_l$ due to risk-$j$}
	\nomenclature[N]{$t_p(x_0)$}{$p$-th quantile of lifetime at normal use stress $x_0$}
	\nomenclature[N]{$x_0$}{standardised normal use stress level}
	\nomenclature[N]{$p$}{quantile level}
	\nomenclature[N]{$s_0$}{normal use stress level (K$^{-1}$)}
	\nomenclature[N]{$B$}{number of Monte Carlo replications}
	\nomenclature[N]{$M_1$}{number of grid points for $\tau$}
	\nomenclature[N]{$C_1(D)$}{Criterion I: preposterior expected posterior variance of $t_p(x_0)$}
	\nomenclature[N]{$C_2(D)$}{Criterion II: preposterior expected posterior variance of $\log( t_p(x_0))$}
	\nomenclature[N]{$D$}{experimental design}
	\nomenclature[N]{$\tau^0$}{optimal stress-change time}
	\nomenclature[N]{$h$}{kernel bandwidth}
	\nomenclature[N]{$K(\cdot)$}{kernel function}
	\nomenclature[N]{$\varphi_{l,j}$}{reparametrised scale parameter for risk $j$ at stress level $x_l$}
	%	\nomenclature[N]{$\pi(\cdot)$}{prior distribution}
	%	\nomenclature[N]{$\pi(\cdot \mid \text{data})$}{posterior distribution}
	\nomenclature[N]{$n_1$}{number of failures before stress-change time $\tau$}
	\nomenclature[A]{NUTS}{No-U-Turn Sampler}
	\nomenclature[A]{MCMC}{Markov chain Monte Carlo}
	\nomenclature[A]{HMC}{Hamiltonian Monte Carlo}
	\printnomenclature[2cm]
	%	\printglossary
	%	\printglossaries
	%	\printnomenclature
	%\contentsname
	
	\onehalfspacing
	
	\section{Introduction}
	\label{sec:intro}
	
	Accelerated life testing (ALT) is a widely used tool 
	for obtaining failure data quickly by exposing items to 
	harsher conditions than they would normally experience, 
	such as increased temperature, humidity, or voltage. 
	ALT is employed in various industries for reasons such 
	as time and cost effectiveness, identifying design flaws 
	before mass production, achieving rigorous standards, 
	and performing predictive analysis on warranty 
	estimations and customer expectations. The two most 
	widely used ALT approaches are constant-stress ALT 
	(CSALT), where a fixed stress level is applied 
	throughout the experiment, and step-stress ALT (SSALT), 
	where the stress is increased at one or more pre-fixed 
	time points. Planning such experiments requires 
	specifying the stress levels, the number of test units, 
	and in SSALT, the stress-change time. Planning an 
	experiment well is important: a poorly chosen 
	stress-change time and stress levels yield imprecise 
	estimates of device reliability at normal use conditions, 
	wasting both time and resources. In a Bayesian 
	framework, the planning problem additionally requires 
	encoding prior engineering knowledge about device 
	behaviour into a probability distribution. Both 
	challenges become considerably more difficult when items 
	are subject to more than one cause of failure and the 
	number of available test units is small.
	
	%	Accelerated life testing (ALT) is widely used tool for examining the 
	%	lifespan and reliability performance of items over a shorter period of 
	%	time than their typical use conditions. This is accomplished by exposing 
	%	the equipment to accelerated stress conditions (such as increased 
	%	temperature, humidity, voltage, and so forth). ALT is employed in 
	%	various industries for a variety of reasons, such as time and cost 
	%	effectiveness, identifying design flaws before mass production, achieving 
	%	rigorous standards, performing predictive analysis on warranty estimations 
	%	and customer expectations, and gaining a competitive advantage in the 
	%	industrial market.
	
	%Constant-stress ALT (CSALT) and step-stress ALT (SSALT) contribute to the accomplishment of one of these primary objectives, which is data	acquisition within a constrained time frame and at a reasonably effective cost. When exposed to various stress conditions, reliability assessment using step-stress accelerated life testing is crucial. 
	For instance, consider a solar lighting device that is prone to two independent 
	failure modes, namely capacitor failure and controller failure 
	\citep{han2014inference}. Since failures at normal operating temperature 
	($293$~K) are unlikely within a reasonable time window, a simple SSALT 
	experiment was conducted by increasing the temperature to $353$~K at a 
	pre-fixed stress change time, in order to obtain accelerated failures. 
	An ideal and potential ALT design/planning provides precise estimations 
	of units' lifetimes under use stress in a timely manner, saving resources 
	and time while maximizing efficiency. However, carrying out ALT tests on 
	small sample data using standard statistical approaches that are often 
	constructed on the basis of asymptotic results makes it difficult to 
	arrive at precise and trustworthy conclusions. In order to efficiently 
	set up SSALT experiments, we describe a Bayesian approach also suited for 
	small sample sizes in this study.
	
	The related studies focused extensively on ALT design, which are ideal 
	for ALT experiments involving large sample sized data. These studies are 
	developed and updated for various kinds of optimal criteria by employing 
	log-location-scale lifetime probability models such as exponential, 
	Weibull, and lognormal distributions. A significant amount of work has 
	been done on designing ALT experiments with a single mode of failure, 
	see \citep{chaloner1992bayesian, ginebra1998minimax, erkanli2000simulation, 
		zhang2005bayesian, zhang2006bayesian, liu2010planning, yuan2011bayesian, 
		yuan2011planning, hamada2015bayesian, you2017self, guan2018bayesian, 
		zhou2020bayesian} and many more citations mentioned in there, which 
	resulted in introducing many criteria minimizing Bayes risk and estimating 
	information matrix (D-, C-, V-optimality), credible intervals, and the 
	pre-posterior variance of quantile and hazard function; 
	cost-constrained optimal designs for constant-stress and 
	step-stress ALT have also been considered 
	\citep{han2015time}. Later, the 
	reference optimality criteria, which relies on Kullback-Leibler divergence, 
	was introduced to build ALT designs \citep{xu2015bayesian, xu2015reference, 
		xu2016planning}. Crucially, the Bayesian planning methods in this 
	literature, even when using MCMC for posterior computation, rely either 
	on a large-sample normal approximation to the posterior or on conjugate 
	prior structures, which are not appropriate in the small-sample regime 
	that motivates the present work.
	
	A comparatively small portion of the studies in this field emphasised on 
	ALT planning using lifetime models with two or more competing hazardous 
	modes. A section of this literature presents the optimal planning for 
	CSALT given multiple competing failure modes, that determines the optimal 
	stress levels and the number of units exposed to those stress levels. To 
	develop optimal designs, these authors used optimal criteria based on the 
	information matrix, its determinant and variance of the estimators of 
	quantile and the hazard function, see \citep{pascual4118444, 
		pascual2008accelerated, fan2021comparison}, whereas 
	\cite{roy2016bayesian} and \cite{roy2018bayesian} attempted to use these criteria 
	using a Bayesian approach while taking prior information into 
	consideration. Specifically, \cite{roy2016bayesian} developed 
	Bayesian D-optimal CSALT plans under competing exponential causes of 
	failure, and \cite{roy2018bayesian} extended this to Weibull component 
	lifetimes for series systems. Also, \cite{Liu5948406} attempted to 
	propose optimal designs under multiple SSALT based on determinant-based 
	optimal criteria for independent Weibull competing risks, 
	and \citet{samanta2019} analysed a Weibull 
	step-stress model under Bayesian framework in the presence of competing risks. As pointed by 
	\cite{roy2016bayesian}, planning methods based on classical approaches 
	that uses maximum likelihood estimation provides locally optimal ALT 
	designs; hence, scenarios where model parameters' values are predicted 
	with uncertainty must be considered. Furthermore it is crucial to design 
	the ALT that are efficient for small sampled experiments as well. 
	However, the existing Bayesian works on competing risks ALT design rely 
	on asymptotic posterior approximations or restrict attention to CSALT, 
	and none of them addresses the prior elicitation problem for the SSALT 
	competing risks model in a principled way. The key challenge is that the 
	natural model parameters $(a_j, b_j, \beta_j)$ are not physically 
	interpretable and are strongly dependent, making it difficult to specify 
	meaningful priors, a problem that has not been solved in the existing 
	literature.
	
	The fundamental purpose of this study is to provide optimal planning 
	strategy for SSALT experiment based on Bayesian guiding principles, 
	regardless of the sample size. Our method takes into consideration the 
	presence of two independent competing failure modes which lead to the 
	failure of an item, providing their lifetimes follow Weibull distributions. 
	We emphasise here on optimizing a criterion that is based on the 
	pre-posterior variance of the $p$-th quantile of the lifetime distribution. 
	A key methodological contribution is a reparametrisation of the SSALT 
	model parameters $(a_j, b_j, \beta_j)$ to a vector 
	$\boldsymbol{\varphi}_j = (t_{0,j}^q, -b_j, \beta_j)$, where $t_{0,j}^q$ is the $q$-th 
	quantile of the failure time distribution due to risk $j$ at use stress. 
	Building on the approach of \citet{zhang2006bayesian} for single-mode 
	SSALT, these transformed parameters are approximately independent and physically interpretable, 
	which enables principled prior elicitation from engineering 
	knowledge at the use stress condition, knowledge that engineers possess 
	naturally and that cannot be readily translated into priors on the 
	original $(a_j, b_j)$ parameters.
	
	We propose an algorithm utilizing No-U-Turn Sampler (NUTS) technique and 
	Monte Carlo simulation that facilitates the best potential Bayesian design. 
	NUTS technique is an extension of Hamiltonian Monte Carlo (HMC) and an 
	alternative to Gibbs sampling. The purpose of this algorithm is to 
	determine a design that maximizes the informative significance of the data 
	gathered during the experiment as well as the prior information that is 
	directly or indirectly related to the model parameters. This is done by 
	minimizing the pre-posterior variance of the $p$-th quantile. 
	Additionally, a comprehensive sensitivity analysis is provided with 
	respect to the quantile probability $p$, the lower stress level $s_1$, 
	the prior hyperparameter choices, and the sample size $n$.
	
	To summarize, this paper aims to provide a thorough 
	framework for structuring SSALT experimentation using a Bayesian approach 
	that works for small sample sizes. In order to illustrate the practical 
	use of the proposed planning, a real-life example of a solar lighting 
	device subject to two competing failure modes \citep{han2014inference} is 
	used for prior elicitation and optimal design. This study differs from 
	previous studies in the following perspectives, and it thus hopes to make 
	a significant contribution to the literature: (1) The proposed design is 
	not locally optimal as it is based on Bayesian methodology, which is 
	highly effective when model-parameters cannot be anticipated with 
	assurance; (2) it does not use any asymptotic results and, therefore the 
	proposed ALT designs are efficient regardless of the sample sizes; 
	(3) a reparametrisation to $\varphi$ parameters enables physically-motivated 
	and approximately independent prior elicitation, which is a novel 
	contribution relative to existing SSALT competing-risks design work; 
	(4) the proposed planning is for two statistically independent Weibull 
	competing risks under simple SSALT which can be extended to 
	multiple stress levels and multiple competing failure modes; (5) this 
	study uses the cumulative exposure model where a log-linear stress-life 
	relationship is used; (6) a comprehensive
	sensitivity analysis is provided with respect to the quantile
	probability $p$, the lower stress level, the prior hyperparameter
	choices, and the sample size.
	
	The remainder of this article is organised as follows. Section~\ref{sec:model} 
	introduces the step-stress competing risks model and the life-test assumptions, 
	including the cumulative exposure model and the log-linear stress-lifetime 
	relationship under the Arrhenius framework. Section~\ref{sec:prob_formulation} 
	formulates the Bayesian planning problem, covering the reparametrised prior 
	specification, the preposterior variance optimality criteria, and the Monte 
	Carlo algorithms used to locate the optimal design. Section~\ref{sec:motivation} describes the solar lighting device dataset that motivates the study and demonstrates 
	the prior elicitation procedure in Subsection~\ref{subsec:prior}. 
	Section~\ref{sec:conv_diag} presents a convergence diagnosis of the NUTS 
	sampler for a representative intermediate lower stress level, establishing 
	the reliability of the posterior estimates before the main simulation study. 
	The comprehensive sensitivity analysis on the optimal designs is carried out 
	in Section~\ref{sec:sa}. Finally, Section~\ref{sec:conclusion} summarises 
	the main findings and outlines directions for future work.
	
	\section{Model Description}
	\label{sec:model}
	
	Suppose that $n$ identical items are put on a simple SSALT experiment 
	and the items are exposed to a stress level $s_1$. Let $\tau$ be a 
	pre-determined time-point where the stress level is changed from $s_1$ 
	to $s_2$. Therefore, items that survived the time-point $\tau$ are 
	continued to be tested under the stress level $s_2$ until a pre-fixed 
	termination time-point $t_c$. Items that are still functioning beyond 
	$t_c$ have their lifetimes censored at $t_c$, and hence $t_c$ is also 
	referred to as the censoring time. The stress variable $s$ is a 
	quantitative measure of the applied stress, such as temperature, 
	voltage, or humidity, depending on the experiment type 
	\citep{nelson1990accelerated, meeker1998statistical}. In the case study 
	presented later in Section~\ref{sec:sa}, temperature is used as the stress 
	variable, where $s_l = 1/T_l$ with $T_l$ denoting the absolute 
	temperature in Kelvin, in accordance with the Arrhenius relationship 
	\citep{meeker1998statistical}.
	
	We assume the following assumptions: \vspace{-\parskip}
	\begin{itemize}
		\setlength{\itemsep}{-\parskip}
		\item  $\tau < t_c$.
		\item  $s_1 < s_2 $.
		\item $\tau, ~ t_c, ~ s_1$ and $s_2$ are pre-fixed before starting the experiment. 
	\end{itemize}  \vspace{-\parskip}
	
	Let $x_l$ be the standardized stress level, then 
	$x_l = (s_l-s_0)/(s_2-s_0),~l = 1, 2$, if $s_0$ denotes the use 
	stress level. Further, define $T_{i,j}(x)$ as the lifetime of the 
	$i$-th item which is exposed to the stress level $x$ and failed due 
	to risk-$j$. Under competing risks, an observed failure lifetime of 
	the item would be $T_i(x) = \min\{ T_{i,1}(x), T_{i,2}(x)\}$, 
	because the item may stop functioning due to two risks under the 
	assumptions. Therefore, we observe $(T_i(x), C_i)$ whenever the item 
	stops functioning during the experiment, where $C_i$ denotes its type 
	of failure that is defined as follows
	\[
	C_i = \begin{cases}
		0, 		& 	~\text{if the item is censored}, \\
		j,  	&	 ~ \text{if the item failed due to risk-}j
	\end{cases}
	\]
	
	Note that $C_i \in \{0,1,2\}$, and for an arbitrary item, it will be 
	denoted by $C$. Now, since it is assumed that the experiment is 
	terminated at time $t_c$, the possible observed lifetime data is 
	given by
	\[ 
	\begin{cases}
		\text{no observation}, & \text{if} ~ n_c = 0 \\ %\text{if} ~ C_i = 0 ~ \forall ~ i = 1 (1) n \\
		(T_{1:n}(x(t)), C_{[1]}), (T_{2:n}(x(t)), C_{[2]}), \dots, \\ (T_{n_c:n}(x(t)), C_{[n_c]}) ,  & \text{if} ~ n_c > 0 % \text{if} ~ C_i \neq 0  ~ \text{for at least one} ~   i  ~\text{among}  ~ i = 1 (1) n  , 
	\end{cases}
	\]
	where
	\begin{align*}
		& 	n_c = ~\text{number of failed items before time} ~ t_c &  \\
		& 	x(t) = ~ x_1 I_{ \{t < \tau\} } + x_2 I_{ \{ t  > \tau \} } &\\
		& 	T_{i:n}(x(t)) = ~ i\text{-th order statistic of the random sample} & \\
		&  ~ ( T_{1}(x(t)), T_{2}(x(t)), \dots, T_{n}(x(t)) )& \\
		& 	C_{[i]} = ~ C\text{-variate paired with the} ~ i\text{-th order statistic} ~ T_{i:n}(x(t)). &
	\end{align*}
	where $I_{\{\cdot\}}$ denotes the indicator function. Note that $x(t)$ indicates a time-variant stress at time 
	$t$ and the $C$-variates are known as concomitants of the order 
	statistics $(T_{1:n}(x(t)), T_{2:n}(x(t)), \dots, T_{n:n}(x(t)))$.
	
	It is assumed that the competing risks are independent and the 
	lifetime of an item that failed due to risk-$j$ at a constant stress 
	$x_l$ follows a Weibull distribution with scale parameter 
	$\theta_j(x_l) > 0$ and shape parameter $\beta_j > 0$, having the following CDF and PDF
	\begin{align}
		\label{eq1}
		F_W(t;\theta_j(x_l),\beta_j) =
		\begin{cases}
			0, 		&	 \text{if} ~ t < 0 \\
			1 - e^{ - \big( \frac{t}{\theta_j(x_l)} \big)^{\beta_j} }, 	&	 \text{if} ~ t > 0
		\end{cases}
	\end{align} 
	and 
	\begin{align}
		\label{eq2}
		f_W(t;\theta_j(x_l),\beta_j) =
		\frac{\beta_j}{\theta_j^{\beta_j}(x_l)}  t^{\beta_j-1} e^{ - \big( \frac{t}{\theta_j(x_l)} \big)^{\beta_j} }   I_{\{t > 0\}},
	\end{align} 
	respectively. The cumulative exposure model (CEM), introduced by 
	\cite{nelson1980accelerated} and thoroughly described by 
	\cite{meeker1998statistical}, is the most widely used framework 
	for analysing step-stress ALT data. Its central assumption is that 
	the remaining lifetime of a test unit depends only on the total 
	accumulated damage up to the current time, regardless of the history 
	of stress application, with the failure time CDF required to be 
	continuous at each stress change point. Alternative frameworks include the tampered failure rate 
	(TFR) model \citep{bhattacharyya1989tampered}, the tampered 
	random variable (TRV) model \citep{degroot1979estimation}, 
	and the failure rate-based model \citep{kateri2015inference}. The CEM is adopted here for its physical interpretability, well-established statistical properties 
	under Weibull and other log-location-scale models 
	\citep{meeker1998statistical}. Under the cumulative exposure model 
	\citep{nelson1980accelerated, meeker1998statistical}, the 
	cumulative effect of the stress exposure for the $j$-th risk at time 
	$t$ is
	\begin{align}
		\label{eq3}
		\psi_{l,j}(t) & =  \left( \frac{t}{\theta_j(x_1)} \right)  I_{\{l=1\}}  + \left(\frac{\tau}{ \theta_j(x_1)} + \frac{t-\tau}{\theta_j(x_2)} \right) I_{\{l=2\}} & \nonumber \\ 
		&   =  \frac{\tilde{t}_{l,j}(t)}{\theta_j(x_l)}  
		%		\psi_{l,j}(t)  = & \begin{cases*}
			%			\frac{t}{\theta_j(x_1)}& \text{for} ~ l = 1\\
			%			\big\{ t - \tau + \frac{\theta_j(x_2)}{\theta_j(x_1)} \tau \big\} \frac{1}{\theta_j(x_2)} & \text{for} ~ l = 2
			%		\end{cases*} & \\
	\end{align} with
	\begin{align}
		\tilde{t}_{l,j}(t)=   \begin{cases}
			t  & \text{for} ~ l = 1\\
			t - \tau + \frac{\theta_j(x_2)}{\theta_j(x_1)} \tau  & \text{for} ~ l = 2. 
		\end{cases} 
	\end{align}
	Therefore, based on the cumulative exposure model, the cause-specific 
	CDF and PDF of the lifetime random variable associated with an item 
	exposed to stress level $l$ that fails due to risk $j$ can be 
	expressed as
	\begin{align*}
		& G_{l,j}(t) = F_W(\tilde{t}_{l,j}(t); \theta_j(x_l), \beta_j),
		~\text{and} & \\
		& g_{l,j}(t) = f_W(\tilde{t}_{l,j}(t); \theta_j(x_l), \beta_j)
		~\text{for}~I(t) = l,~C = j,~j \in \{1, 2\}, &
	\end{align*}
	respectively. Here $I(t)$ denotes the index of the stress level at time-point $t$ and $j'$ represents the complement of $j$. Since only the minimum $T = \min\{T_1, T_2\}$ is observed, the overall CDF of a lifetime random variable associated with an item that may disfunction possibly due to two independent 
	risks under simple SSALT experiment is
	\begin{align}
		\label{eq4}
		F(t) = 1 - \prod_{j=1}^{2}\Big(1 - G_{l,j}(t)\Big), \quad
		l = \begin{cases} 1 & \text{if}~t < \tau \\ 
			2 & \text{if}~t \geq \tau. \end{cases}
	\end{align}
	Furthermore, let $C_i$ denote the indicator variable for the cause 
	of failure. Then, the joint distribution of $(T_i, C_i)$ is given by
	\begin{align}
		\label{eq_joint}
		f_{T,C}(t, j) = & g_{l,j}(t)\prod_{\substack{j'=1\\j'\neq j}}^{2}
		\Big(1 - G_{l,j'}(t)\Big) & \nonumber \\
		= & g_{l,j}(t)\Big(1 - G_{l,j'}(t)\Big),
		\quad j \in \{1,2\}, &
	\end{align}
	which gives the likelihood contribution of an item that fails at 
	time $t$ due to risk $j$ under stress level $l$. For a censored 
	item, the contribution is $\prod_{j=1}^{2}(1 - G_{2,j}(t_c))$.
	\begin{align*}
		L(\underline{\boldsymbol{\theta}}|\text{data}) \propto ~
		& \prod_{i =1}^{n}   \left\{   \prod_{l =1}^{2} \prod_{j =1}^{2}  {\bigg[ 	g_{l,j}(t_{i:n}) (1 - G_{l,j'}(t_{i:n}))  \bigg]}^{I_{\{ C_i=j, I(t_{i:n})=l \}}} \right\}  {\bigg[  \prod_{j =1}^{2} \Big( 1 - G_{2,j}(t_c)	\Big) \bigg]}^{I_{\{ C_i=0 \}}}    & \\ 
		\propto ~ & \prod_{i =1}^{n}   \left\{   \prod_{l =1}^{2} \prod_{j =1}^{2}  {\bigg[ 	f_W( \tilde{t}_{l,j}(t_{i:n}); \theta_j(x_l), \beta_j )  \Big(1 - F_W( \tilde{t}_{l,j'}(t_{i:n}); \theta_{j'}(x_l), \beta_{j'} )\Big)   \bigg]}^{I_{\{ C_i=j, I(t_{i:n})=l \}}} \right\} &\\
		& \times  {\bigg[  \prod_{j =1}^{2} \left( 1 -  F_W(\tilde{t}_{2,j}(t_c) ; \theta_j(x_2), \beta_j)	\right) \bigg]}^{I_{\{ C_i=0 \}}}    & \\ 
		\propto ~ &  {\bigg[   e^{- \sum_{j=1}^{2} \psi_{2,j}^{\beta_j}(t_{c}) } \bigg]}^{n-n_c}   \prod_{i =1}^{n_c}   \left\{  \prod_{l =1}^{2} \prod_{j =1}^{2}  {\bigg[ 	\frac{\beta_j}{\theta_j(x_l)} \psi^{\beta_j - 1}_{l,j}(t_{i:n}) e^{ - \sum_{j=1}^{2}    \psi^{\beta_j}_{l,j}(t_{i:n})   }   \bigg]}^{I_{\{ C_i=j, I(t_{i:n})=l \}}} \right\} &
		%		\propto ~ &      e^{- (n-n_c)\sum_{j=1}^{2} \psi_{2,j}^{\beta_j}(t_{c}) }    \prod_{j =1}^{2}  \prod_{l =1}^{2}     \Bigg\{ 	\frac{\beta_j^{n_{l,j}}}{\theta_j^{n_{l,j}}(x_l)} \prod_{i =1}^{n_c}   {\bigg[ \psi^{\beta_j - 1}_{l,j}(t_{i:n}) e^{ - \sum_{j=1}^{2}    \psi^{\beta_j}_{l,j}(t_{i:n}) } \bigg]}^{I_{\{ C=j, I(t_{i:n})=l \}}}   \Bigg\}.   &
	\end{align*}
	%		\hrulefill
	%		\vspace*{4pt}
	%	\end{strip}

where $n_c$ indicates the number of failures occurred before the 
censoring time. For computational 
implementation in the Stan, the log-likelihood is used directly, given by
\begin{align}
	\label{eq_loglik}
	\ell(\underline{\boldsymbol{\theta}}|\text{data}) & = 
	\text{const} - (n-n_c) \sum_{j=1}^{2} \psi_{2,j}^{\beta_j}(t_{c}) & \nonumber \\
	& +  \sum_{i =1}^{n_c}  \sum_{l =1}^{2} \sum_{j =1}^{2}  I_{\{ C_i=j, I(t_{i:n})=l \}} & \nonumber \\
	& \times 	\bigg[ 	\log(\beta_j) - \log(\theta_j(x_l)) + (\beta_j-1)\log(\psi_{l,j}(t_{i:n})) & \nonumber 	\\
	&	- \sum_{m=1}^{2}\psi_{l,m}^{\beta_m}(t_{i:n})  \bigg] &
\end{align}

In this study, it is assumed that the scale parameter of the Weibull lifetime model is log-linearly dependent on the stress level with the 
following relationship
\begin{align}
	\label{eq5}
	\log(\theta_j(x_l)) = a_j + b_j x_l,~j = 1, 2~\text{and}
	~l = 1, 2,
\end{align}
where $a_j \in \mathbb{R}$ is the intercept and $b_j$ is the slope parameter for 
risk-$j$ in the lifetime-stress relationship. Note that $b_j < 0$, since higher stress levels result in shorter 
lifetimes. The log-linear form in \eqref{eq5} arises naturally from the 
Arrhenius acceleration model, which is the most widely used model 
for thermally activated failure mechanisms \citep{meeker1998statistical, 
	nelson1990accelerated}. The Arrhenius model postulates that the rate 
of a degradation reaction at absolute temperature $T$ (Kelvin) is 
proportional to $\exp(-E_a / kT)$, where $E_a > 0$ is the 
activation energy of the failure mechanism (eV) and 
$k = 8.6171 \times 10^{-5}$~eV/K is Boltzmann's constant. Since the Weibull scale parameter $\theta$ decreases with increasing degradation rate (i.e., faster degradation implies shorter lifetime), 
we have $\theta(T) \propto \exp(E_a/kT)$, or equivalently,
\[
\log (\theta(T)) = a^* + \frac{E_a}{k} \cdot \frac{1}{T},
\]
for some intercept $a^*$. The natural stress variable is therefore 
$s = 1/T$ (inverse Kelvin). Since each failure mode governs a distinct degradation 
mechanism, each carries its own activation energy $E_{a,j}$, and after standardising to 
$x_l = (s_l - s_0)/(s_2 - s_0)$, the log-linear form 
$\log(\theta_j(x_l)) = a_j + b_j x_l$ in \eqref{eq5} follows 
directly, with 
\[
a_j = a^* + \frac{E_{a,j}}{k} \cdot \frac{1}{T_0} = \log(\theta_j(T_0))
\] 
and 
\[ 
b_j = \frac{E_{a,j}}{k} \cdot \left( \frac{1}{T_2} - \frac{1}{T_0} \right) < 0  ~(\text{negative because } T_2 > T_0).
\]
Furthermore, the shape parameters $\beta_j$ are assumed to be constant 
across stress levels \citep{meeker1998statistical}. 

Now we consider the quantity of interest which is the $p$-th quantile of the lifetime 
distribution that satisfies
\begin{align*}
	P(T(x_0) \leq t_p(x_0)) = p,
\end{align*}
which is equivalent to
\begin{align}
	\label{eq6}
	(t_p(x_0) e^{-a_1})^{\beta_1} +
	(t_p(x_0) e^{-a_2})^{\beta_2} = -\log(1-p).
\end{align}
$t_p(x_0)$ is a solution of the non-linear equation in \eqref{eq6}, 
obtained numerically using the Newton-Raphson method.

In reliability engineering, lower-tail quantiles (small $p$, 
such as $p = 0.10$) are of particular practical concern, 
governing warranty periods and safety margins. The optimality 
criteria in Section~\ref{sec:prob_formulation} are built 
directly around the posterior variance of $t_p(x_0)$, so 
that the optimal design minimises posterior uncertainty about 
the true quantile value at use conditions.

\section{Problem Formulation on Bayesian Planning}
\label{sec:prob_formulation}
A prior information of the SSALT model parameters is essential when developing a Bayesian design.  Therefore, we evaluate a number of potential choices in Subsection \ref{subsec:prior_selection} for choosing the priors before presenting the prior assumptions under this study.
\subsection{Prior Selection}
\label{subsec:prior_selection}
The most commonly used and straightforward approach would be to use the independence of the parameters $(a_1,b_1,\beta_1, a_2,b_2,\beta_2)$ to assume that
\begin{align*}
	f_{\underline{\boldsymbol{\theta}}}(a_1,b_1,\beta_1, a_2,b_2,\beta_2) = &  f_{a_1}(a_1) f_{b_1}(b_1) f_{\beta_1}(\beta_1)  f_{a_2}(a_2) f_{b_2}(b_2) f_{\beta_2}(\beta_2) &
\end{align*} 
is the joint prior density. This assumption has certain shortcomings: (1) The independence of the parameters $a_j, ~ b_j$, and $\beta_j$ is difficult to defend; (2) It is hard to elicit prior distributions for the intercept $a_j$ and the slope parameter $b_j$ for any particular risk $j$. As a result, we typically lack direct prior knowledge of the SSALT parameters in real situations. Furthermore, it is difficult to justify the statistical independence of the subjective priors, even in the case that they are known separately.  To overcome this hurdle, engineers often come up with their subjective beliefs concerning the unknown model parameters using alternative lifetime characteristics that are more easily specified in real-life applications. Therefore,  In most practices, it is easier to acquire information on the functions of parameters such as various quantiles at usage stress levels and median lifetime at different stress levels, etc..

\cite{singpurwalla2006reliability} (Section 5.2.2 page 128) suggested to consider independent priors for the shape parameter and the median of the distribution at different stresses that are not necessarily to be the testing stress levels . Statistical Independence of the median lifetime of the Weibull distribution and its shape parameter is justified by \cite{singpurwalla1988interactive}. According to the author, subject matter specialists are able to conceptualize and assess median lifetimes without any encumbrance from their assessment about aging. These prior choices are implemented by \cite{yuan2011bayesian,yuan2011planning, guan2018bayesian} for single-cause SSALT. Another alternative is presented by \cite{zhang2006bayesian} for single-cause SSALT where independent priors are taken for the $q$-th quantile at use stress level, shape and the slope parameters. According to \cite{meeker1998statistical} (Section 1.4.3), a small quantile of a lifetime distribution is approximately independent of its shape parameter for heavily censored experiments. Further, \cite{xu2015reference, xu2016planning} assumed priors on the $q$-th quantile at use stress level, the slope parameter and the median lifetime of the failure distribution at highest stress level. 

This study based on the competing risks plans to employ the approach suggested by \cite{zhang2006bayesian} and implemented by \cite{roy2018bayesian}. Let us denote the $\varphi_{j}=(\varphi_{1j},\varphi_{2j},\varphi_{3j})$ for $j=1,2$, where
\begin{align}
	\label{eq7}
	\begin{cases}
		\varphi_{1j}=  t_{0,j}^q = e^{ a_j} (-\log(1-q))^{1/\beta_j} & \\
		\varphi_{2j}= -b_j & \\
		\varphi_{3j}= \beta_j. & 
	\end{cases}
\end{align} Here the quantity $ t_{0,j}^q$ denotes the $q$-th quantile of failure time distribution due to risk-$j$ at usage stress condition.
% $ \varphi_{1j}=  t_{0,j}^q = e^{ a_j} (-\log(1-q))^{1/\beta_j}$ which denotes the $q$-th quantile of failure time distribution due to risk-$j$ at usage stress condition, $ \varphi_{2j}= -b_j$ and $ \varphi_{3j}= \beta_j$. 
Therefore, we have $\varphi_{j}=(t_{0,j}^q , -b_j, \beta_j)$. These three quantities in the vector $\boldsymbol{\varphi}_j$ are always positive and so, any PDF with positive support such as, lognormal, uniform, gamma, etc., may be suitable as their prior densities. In this study, we assume the $\varphi_{ij'}$s have independent gamma PDFs.  Now this prior knowledge can be directed further to obtain the joint prior information for the SSALT model parameters $a_1,b_1,\beta_1, a_2,b_2$ and $\beta_2$. Suppose $f_{G_{l,j}}(.)$ denotes the gamma PDF for the random quantity $\varphi_{ij'}$, then using the multivaraite transformation of random variables, the joint PDF for the parameter $\underline{\boldsymbol{\theta}}$ is given by
\begin{align}
	\label{eq8}
	f_{\underline{\boldsymbol{\theta}}}(a_1,b_1,\beta_1, a_2,b_2,\beta_2) = & \prod_{j=1}^{2} e^{a_j} (-\log(1-q))^{1/\beta_j} &  \nonumber  \\ 
	&  \times  f_{G_{1,j}}( e^{a_j} (-\log(1-q))^{1/\beta_j})  f_{G_{2,j}}(-b_j)  f_{G_{3,j}}(\beta_j), & 
\end{align} where $a_j \in \mathbb{R}, b_j < 0, \beta_j > 0$ for $j = 1, 2.$ Note that the expression of the gamma PDF priors $ f_{G_{l,j}}( .)$ for a given $l$ and $j$ involves the hyper parameters. The choices of the hyper parameters depends on the choice of the data. We will have a detailed discussion on the prior choices while doing the simulation study in Subsection \ref{subsec:prior}.
Although Stan operates entirely in the $\boldsymbol{\varphi}_j$ 
parameterisation, Equation~\eqref{eq8} is included to make 
explicit the joint prior induced on the original parameters 
$(a_j, b_j, \beta_j)$. The non-standard form of the induced 
prior in~\eqref{eq8} underscores why direct prior elicitation 
on $(a_j, b_j, \beta_j)$ is impractical.
\subsection{Bayesian Criterion}  
\label{subsec:bayesian_criteria}
%$\bullet$ What criterion is used and how to get an optimal design?  \\ 
%$\bullet$ What is optimal design?  \\ 
%$\bullet$ Notations used for one decision variable design and two decision variable design.
An ALT experiment is frequently designed to estimate a quantile of the life distribution, generally near its lower tail, under certain usage conditions.  Because of this, it makes sense to develop a Bayesian ALT design based on the accuracy of guessing $t_p(x_0)$. In this case, we define the first criterion function as the posterior variance of $t_p(x_0)$. On the other hand, the posterior variance depends on the observed data and is conditioned on a specific plan ALT design denoted by $D$ (which includes the accelerating stress levels and their changing time points). The goal of an ideal strategy must be to increase its usefulness. Therefore, in order to formulate a Bayesian test planning criterion, we compute the preposterior expectation of the posterior variance over the entire marginal distribution of $t$. Thus, our first criterion denoted by $C_1(D)$ becomes
\begin{align}
	\label{eq9}
	C_1(D) = \text{E}_{\text{data}}(\text{Var}_{\underline{\boldsymbol{\theta}}|\text{data}}(t_p(x_0)))
\end{align}
Since the quantity $t_p(x_0)$ is positive, our second criterion function is based on posterior variance of $\log(t_p(x_0))$ and it is given by
\begin{align}
	\label{eq10}
	C_2(D) = \text{E}_{\text{data}}(\text{Var}_{\underline{\boldsymbol{\theta}}|\text{data}}(\log(t_p(x_0))))
\end{align}

$C_1(D)$ targets precision in absolute time units, which 
is natural for warranty and maintenance planning. $C_2(D)$ 
operates on the log scale and is scale-free, making it 
more suitable for comparing designs across different 
quantile levels where $t_p(x_0)$ varies considerably 
in magnitude. The two criteria need not yield the same optimal design. The above criteria $C_1(D)$ and $C_2(D)$ are similar to a criterion function generated from a quadratic loss function used by \cite{erkanli2000simulation}. To obtain a Bayesian simple SSALT design, the proposed criterion function in Equations \eqref{eq9} and \eqref{eq10} are minimized over all possible choices of simple SSALT designs. A Simple SSALT design is denoted by $D$ and it represents a bivariate quantity that include two decision variables $s_1$ and $\tau$. An optimal design is the one that has the minimum preposterior variance among all the feasible designs. In this article, we consider two cases when providing optimal design: (1) one-variable optimal design: when the lower stress level is fixed and optimization involves finding  its change time point $\tau$; (2) two-variable optimal design: optimization involves finding two decision variables lower stress level $s_1$ and its change time point $\tau$. Therefore, under two-variable optimal design problem, we have
\begin{align*}
	C_1^0(D^0) = \mathop{\min}_{D}  C_1(D) ~ \text{and} ~ C_2^0(D^0) = \mathop{\min}_{D}  C_2(D)
\end{align*} where $D^0 = (s_1^0,\tau^0)$ is optimal design, whereas 
\begin{align*}
	C_1^0(\tau^0) = \mathop{\min}_{\tau}  C_1(\tau) ~ \text{and} ~ C_2^0(\tau^0) = \mathop{\min}_{\tau}  C_2(\tau)
\end{align*} for a pre-fixed lower stress $s_1 $under one-variable optimal design problem with $\tau^0$ indicating the optimal design. The notations $C_1^0$ and $C_2^0$ are used to indicate the corresponding optimal criterion values.
Or, in other ways
\begin{align*}
	D^0= \mathop{\argmin}_{D}  C_1(D) ~ \text{and} ~ D^0 = \mathop{\argmin}_{D}  C_2(D)
\end{align*} and
\begin{align*}
	\tau^0 = \mathop{\argmin}_{\tau}  C_1(\tau) ~ \text{and} ~ \tau^0 = \mathop{\argmin}_{\tau}  C_2(\tau)
\end{align*}
\subsection{Optimization Algorithms} 
\label{subsec_opt_algo}
%Key points: Monte-carlo, HMC and No-U-Turn, Gibbs sampling (references), smoothed surface technique, steps for algorithm, Stan package
Since the optimality criteria are based on the pre-posterior variance estimates of the quantiles. We need to obtain the posterior variance estimates for the quantile of the lifetime distribution of the items. Direct computation of the posterior estimators is not possible due to the intricacy of the joint posterior density under this study. For approximation, numerical approaches are therefore required. In this work, we simulate posterior densities using the NUTS, an extension of the HMC approach. From these densities, NUTS extracts large representative samples, which are subsequently utilised to compute approximate posterior estimators for the unknown parameters.

By utilizing Hamiltonian dynamics, the HMC method, a generalization of the Metropolis algorithm, improves target distribution exploration. It is highly sensitive with respect to the tuning-parameters step size and the number of steps. In order to improve performance, NUTS, an adaptive variant of HMC, automatically modifies these tuning-parameters without requiring user input. With the release of Stan, a free and open-source \citep{gelman2015stan,carpenter2017stan}, that is a probabilistic programming language written in C++, NUTS has become much more popular. Stan uses NUTS as its primary algorithm for inferences, which enables Bayesian inference to be implemented in a variety of programming languages including R. Numerous fields have widely adopted it because of its speed, ease of use, and frequent updates. Here we implement the `Stan' using R interface.

The variance estimates of $t_p(x_0)$ and $\log(t_p(x_0))$ based on the posterior distribution are calculated using the \texttt{cmdstanr} package in R. There are certain steps to achieve the posterior estimates using `Stan' in R that can be described as follows:
\begin{enumerate}
	\item Begin by importing and preparing the data in R. This initial step involves data importation and cleaning procedures to ensure the data set is ready for analysis. 
	\item Define the statistical model in Stan. This crucial step involves articulating the statistical model using the Stan language, which can be done either in a separate file or as a string within R. The Stan model declares the reparametrised quantities $\varphi_{ij}$, recovers the original model parameters and computes $t_p(x_0)$, and evaluates the log-likelihood in~\eqref{eq_loglik}.
	%	Here, we specify the model structure in Listing \ref{lst:lst1} as follows: 
	%	\lstinputlisting[style=mystyle, language=C++, caption={Stan Model}, label={lst:lst1}]{C:/Users/Kiran/.cmdstan/cmdstan-2.35.0-rc3/examples/bernoulli/bernoulli_KP.stan} 
	\item Compile the Stan model. Using the \texttt{cmdstan\_model()} function in R, we compile the Stan model.
	\item Transmit the data and initial values to the Stan model and execute the model. We pass the cleaned data and any relevant parameters to the Stan model for the analysis using the \texttt{sample$()$} function. This step initiates the execution of the model, generating posterior distributions for the parameters of interest.
	\item Extract posterior draws and conduct parameter inference using R. Following the model execution in Stan as mentioned in the above point, we utilize R to extract draws from the posterior distributions. With these draws, we perform optimization of the target criterion functions.
\end{enumerate}
%\begin{algorithm}[tbp]
%	\caption{: ~ ~ Steps used for optimization of the criterion functions.}
%	\label{algo_1}
%	\begin{enumerate}[Step 1. ] 
	%			\item Define a Stan model. \vspace{-1em}
	%			\item Compile the Stan model.\vspace{-1em}
	%			\item generate random initial values for the $\varphi_{ij'}$s from gamma distribution. \vspace{-1em}
	%			\item provide the needful  data values. \vspace{-1em}
	%			\item use these initial values and data to get the posterior variance estimates obtained using the sampling method for parameters $t_p$ and $\log(t_p)$. 					
	%	\end{enumerate}
%\end{algorithm} 

%We will use the following two terminologies `one-variable optimal design' and `two-variable optimal design' in the remaining article. When the lower stress $s_1$ is known and the criteria are minimized over the stress changing point $\tau$ to find the optimal $\tau^0$, it is termed as one-variable optimal design. Similarly, the two-variable optimal design can be defined when both $s_1$ and $\tau$ are unknown and criteria are minimized with respect to the bivariate quantity $D$. Recall that we denoted a design by a bivariate quantity $D$. Now, we will provide the algorithms that are used to determine these optimal designs.

The algorithms for determining the one-variable and two-variable optimal designs, as defined in 
Section~\ref{subsec:bayesian_criteria}, are described below.
\subsubsection{One-variable optimal design}
Algorithm \ref{algo_1} is used to find the one-variable optimal design for a given stress level $s_1$. 
\setcounter{algorithm}{0}
\begin{algorithm}[tbp]
	\caption{: ~ Steps used to determine one-variable 
		optimal design.}
	\label{algo_1}
	\begin{enumerate}
		\item Obtain $M_1$ equally spaced grid points 
		$\tau_1, \tau_2, \ldots, \tau_{M_1}$ for the 
		stress-change time $\tau$ over the interval 
		$(\tau_l, \tau_u)$.
		\item For each grid point $\tau_k$, 
		$k = 1, \ldots, M_1$, simulate $B$ random 
		samples from the SSALT experiment under 
		competing risks, with pre-fixed values of $n$ 
		and the model parameters.
		\item For each simulated sample, obtain the 
		posterior variance estimates of $t_p(x_0)$ 
		and $\log(t_p(x_0))$ using the 
		\texttt{cmdstanr} package in R.
		\item Average the posterior variance estimates 
		obtained in Step~3 over all $B$ simulated 
		samples and assign these averages as 
		$C_1(\tau_k)$ and $C_2(\tau_k)$.
		\item Repeat Steps~2 - 4 for each 
		$k \in \{1, 2, \ldots, M_1\}$ to obtain the 
		raw criterion values $C_1(\tau_k)$ and 
		$C_2(\tau_k)$ over the full $\tau$ grid.
		\item Apply the Gaussian kernel smoother 
		in~\eqref{eq11} to the raw criterion curves 
		$\{C_1(\tau_k)\}$ and $\{C_2(\tau_k)\}$ 
		obtained in Step~5, to obtain smooth 
		approximations $C_1(\tau)$ and $C_2(\tau)$ 
		for any $\tau \in (\tau_l, \tau_u)$.
		\item Using a fine grid of $500$ equally 
		spaced points over $(\tau_l, \tau_u)$, locate 
		the minimisers of the smoothed functions 
		$C_1(\tau)$ and $C_2(\tau)$ obtained in 
		Step~6. These minimisers constitute the 
		one-variable optimal designs $\tau^0$ for 
		Criterion I and Criterion II respectively.
	\end{enumerate}
\end{algorithm}

In Step $7$ of Algorithm \ref{algo_1}, a smooth curve is mentioned and 
it is obtained using a kernel smooth technique to get an 
estimate of $\tau^0$. Standard Gaussian kernels are used 
to get the smooth curve for criterion functions over the 
entire range of the design variable $\tau$ and it is 
implemented on the $M_1$ criterion function values that 
we obtained in Step 6. The bandwidth is set equal to the 
grid spacing, that is, $h = (\tau_u - \tau_l)/(M_1 - 1)$. The 
smooth function's value at a point $\tau$, which is an 
approximation of the original criterion functions, is 
obtained as follows
\begin{align}
	\label{eq11}
	C_1(\tau) \approx \frac{\sum\limits_{i=1}^{M_1} K(\frac{\tau-\tau_i}{h})C_1(\tau_i)}{\sum\limits_{i=1}^{M_1} K(\frac{\tau-\tau_i}{h})}
\end{align}
where $K(\cdot)$ denotes the standard normal density. The 
formula in Equation~\eqref{eq11} is also implemented on 
function $C_2(\tau)$. In Step 8, a fine grid of $500$ 
equally spaced points over $(\tau_l, \tau_u)$ is used to 
locate the optimal $\tau^0$ on the smoothed curve. Future 
work may explore alternative kernel functions, such as the 
Epanechnikov or Cauchy kernel, for smoothing the criterion 
surface.

%In Step $7$ of Algorithm \ref{algo_1}, a smooth curve is mentioned and it is obtained using a kernel smooth technique to get an estimate of $\tau^0$. Standard Gaussian kernels are used to get the smooth curve for criterion functions over the entire range of the design variable $\tau$ and it is implemented on the $M_1$ criterion function values that we obtained in Step $6$. The smooth function's value at a point $\tau$, which is an approximation of the original criterion functions, is obtained as follows
%	\begin{align}
	%	\label{eq11}
	%	C_1(\tau) \approx \frac{\sum_{i=1}^{M_1} K(\frac{\tau-\tau_i}{h})C_1(\tau_i)}{\sum_{i=1}^{M_1} K(\frac{\tau-\tau_i}{h})}
	%\end{align} The formula in Equation \eqref{eq11} is also implemented on function $C_2(\tau)$. In this study, we take $K(~)$ to be the standard normal densities. Although, standard Cauchy densities may be used as an alternative choice. 
	
	The boundary points are avoided in the proposed algorithms and it can be justified by the following points: (1) convergence is usually a problem on boundary points (discussed later in Subsection \ref{sec:conv_diag}); (2) applying higher stress in the beginning of the experiment may cause severe damage to the units, hence avoided; (3) it does not make much sense to change the stress level at a point that is very close to the termination period.
	\subsubsection{Two-variable optimal design}
	The two-variable optimal design is determined by 
	optimising the criterion functions jointly over the 
	lower stress level $s_1$ and the stress-change time 
	$\tau$. The complete methodology is described by 
	Algorithm~\ref{algo_2}.
	
	\begin{algorithm}[tbp]
		\caption{: ~ Steps used to determine two-variable 
			optimal design.}
		\label{algo_2}
		\begin{enumerate}
			\item Obtain $M_2$ equally spaced stress points 
			in the interval $(s_{1,l}, s_{1,u})$ for the 
			lower stress variable $s_1$.
			\item For each stress point $s_{1,i}$, 
			$i = 1, \ldots, M_2$, follow Steps~1-6 of 
			Algorithm~\ref{algo_1} for all $\tau_j$, 
			$j = 1, \ldots, M_1$, to build the raw 
			criterion surface $\{C_k(x_{1,i}, \tau_j)\}$ 
			over the full two-dimensional design grid, 
			for $k = 1, 2$.
			\item Apply the two-dimensional kernel smoothing 
			in~\eqref{eq_smooth2d} jointly to the raw 
			criterion surface $\{C_k(x_{1,i}, \tau_j)\}_{i,j}$, 
			to find the smooth criterion surface over $(\tau_l, \tau_u) \times 
			(x_{1,l}, x_{1,u})$, for $k = 1, 2$.
			\item The point $(s_1^0, \tau^0)$ at which the 
			smoothed surface attains its minimum value 
			constitutes the two-variable optimal design 
			$D^0$, separately for each criterion $k = 1, 2$.
		\end{enumerate}
	\end{algorithm}
	
	The two-dimensional kernel smoother in Step~3 of 
	Algorithm~\ref{algo_2} extends the one-variable 
	smoothing in~\eqref{eq11} to the joint design space 
	$(x_1, \tau)$. The smoothed criterion value at a 
	point $(x_1, \tau)$ is obtained as
	\begin{align}
		\label{eq_smooth2d}
		C_k(x_1, \tau) \approx
		\frac{ \sum\limits_{i=1}^{M_2} \sum\limits_{j=1}^{M_1} 
			K \left(\frac{x_1 - x_{1,i}}{h_{x_1}}\right)
			K \left(\frac{\tau - \tau_j}{h_\tau}\right)
			C_k(x_{1,i}, \tau_j)}
		{\sum\limits_{i=1}^{M_2}\sum\limits_{j=1}^{M_1} 
			K \left(\frac{x_1 - x_{1,i}}{h_{x_1}}\right)
			K \left(\frac{\tau - \tau_j}{h_\tau}\right)},
	\end{align}
	where the approximation is applied separately to 
	$C_1(x_1, \tau)$ and $C_2(x_1, \tau)$, $K(\cdot)$ 
	denotes the standard normal density as 
	in~\eqref{eq11}, and $h_\tau = (\tau_u - \tau_l)/
	(M_1 - 1)$ and $h_{x_1} = (x_{1,u} - x_{1,l})/
	(M_2 - 1)$ are the bandwidths in the $\tau$ and 
	$x_1$ directions respectively. The smoothed surface 
	is evaluated on a fine grid of $100 \times 50$ equally 
	spaced points over $(\tau_l, \tau_u) \times 
	(x_{1,l}, x_{1,u})$, and the two-variable optimal 
	design $D^0 = (s_1^0, \tau^0)$ is identified as the 
	joint minimiser of~\eqref{eq_smooth2d}.
	
	%Note that the resulted optimal designs might be different for both criteria into consideration.
	
	% ============================================================
	% Section: Motivating Example and Prior Elicitation
	% Based on Han & Kundu (2014) solar lighting device data
	% To be inserted in main paper before the Simulation/SA section
	% ============================================================
	
	\section{Motivating Example and Prior Elicitation}
	\label{sec:motivation}
	
	The proposed Bayesian design methodology requires the specification of prior
	distributions for the reparametrised model parameters
	$\boldsymbol{\varphi}_j = (t_{0,j}^q,\, -b_j,\, \beta_j)$, $j = 1, 2$.
	In this section we describe the real dataset from which the prior
	hyperparameters are elicited, verify the adequacy of the assumed model for
	that dataset, and detail the elicitation procedure.
	It is important to emphasise that this dataset serves as a
	\emph{separate preliminary dataset}, independent of any future experiment
	to be designed. The prior information is therefore drawn from external
	engineering knowledge and previously observed data, not from the experiment
	whose design is being optimised. This is a standard and legitimate form of
	informative prior elicitation in Bayesian analysis \citep{gelman2013bayesian},
	and it is precisely the situation in which the proposed framework is intended
	to be applied.
	
	\subsection{Dataset Description}
	\label{subsec:data}
	
	We consider the step-stress ALT dataset reported in Table~7 of
	\cite{han2014inference} for a solar lighting device subject to two
	independent competing failure modes: capacitor failure (cause~1) and
	controller failure (cause~2). The stress variable is temperature.
	A simple SSALT experiment was conducted on $n = 35$ identical units.
	Units were initially tested at the lower stress level $s_1 = 293$~K
	(the use temperature condition) and the stress was increased to
	$s_2 = 353$~K at the pre-fixed stress change time $\tau = 5$ (in
	units of $100$ hours). The experiment was terminated at $t_c = 6$
	(i.e., $600$ hours) under Type-I censoring. Of the $n = 35$ units,
	$31$ failures were observed and $4$ units were right-censored at $t_c$.
	Each observation consists of a failure time, the stress level at which
	failure occurred, and the associated cause of failure.
	The complete dataset is reproduced in Table~\ref{table:han_kundu_data}.
	
	\begin{table}[tbp]
		\centering
		\caption{Step-stress ALT data from \cite{han2014inference} for the solar lighting device. Each entry is (failure time in 100 hours, cause of failure).
			Cause~1: capacitor failure; Cause~2: controller failure.}
		\label{table:han_kundu_data}
		\renewcommand{\arraystretch}{1.15}
		\begin{tabular}{cc}
			\hline
			\multicolumn{2}{c}{Stress level} \\
			293~K & 353~K \\
			\hline
			$(0.140,\,1)$ & $(5.002,\,1)$ \\
			$(0.783,\,2)$ & $(5.022,\,2)$ \\
			$(1.324,\,2)$ & $(5.082,\,2)$ \\
			$(1.582,\,1)$ & $(5.112,\,1)$ \\
			$(1.716,\,2)$ & $(5.147,\,1)$ \\
			$(1.794,\,2)$ & $(5.238,\,1)$ \\
			$(1.883,\,2)$ & $(5.244,\,1)$ \\
			$(2.293,\,2)$ & $(5.247,\,1)$ \\
			$(2.660,\,2)$ & $(5.305,\,1)$ \\
			$(2.674,\,2)$ & $(5.337,\,2)$ \\
			$(2.725,\,2)$ & $(5.407,\,1)$ \\
			$(3.085,\,2)$ & $(5.408,\,2)$ \\
			$(3.924,\,2)$ & $(5.445,\,1)$ \\
			$(4.396,\,2)$ & $(5.483,\,1)$ \\
			$(4.612,\,1)$ & $(5.717,\,2)$ \\
			$(4.892,\,2)$ & \\
			\hline
		\end{tabular}
	\end{table}
	
	\subsection{Model Fitting and Goodness-of-Fit}
	\label{subsec:model_fitting}
	
	The step-stress competing risks Weibull model described in Section~\ref{sec:model} 
	was fitted to the dataset by maximum likelihood estimation under the cumulative exposure 
	model. The MLEs of the model parameters $(a_j, b_j, \beta_j)$ for $j = 1, 2$ 
	are reported in Table~\ref{table:mle_par}. The negative values of $\hat{b}_j$ for 
	both causes confirm the expected acceleration effect: lifetime decreases as 
	temperature increases. The shape parameter estimates $\hat{\beta}_1 = 0.769$ 
	and $\hat{\beta}_2 = 1.532$ indicate decreasing and increasing hazard rates 
	for cause~1 and cause~2, respectively, pointing to qualitatively different 
	failure mechanisms for the two competing risks. The observed information matrix 
	at the MLE has a block-diagonal structure corresponding to the two independent 
	causes of failure, consistent with the assumed 
	independence of the competing risks.
	
	\begin{table}[H]
		\centering
		\caption{Maximum likelihood estimates of the step-stress
			competing risks Weibull model parameters under cumulative exposure model.}
		\label{table:mle_par}
		\renewcommand{\arraystretch}{1.2}
		\begin{tabular}{lccc}
			\hline
			Cause & $\hat{a}_j$ & $\hat{b}_j$ & $\hat{\beta}_j$ \\
			\hline
			Cause~1 (capacitor)   & $4.5064$ & $-4.7131$ & $0.7692$ \\
			Cause~2 (controller)  & $2.0410$ & $-1.2277$ & $1.5321$ \\
			\hline
		\end{tabular}
	\end{table}
	
	Goodness-of-fit was assessed using empirical distribution function~(EDF)-based 
	statistics under Type-I censoring. Since the asymptotic null distributions of EDF statistics are not valid 
	when parameters are estimated from the data and observations are censored 
	\citep{stephens1974}, $p$-values were obtained via a parametric bootstrap. Bootstrap samples were generated from the fitted model under 
	the same step-stress structure and censoring, and the 
	KS and CvM statistics were recomputed for each of the $1000$ 
	replicates.
	
\begin{figure}[H]
		\centering
		\includegraphics[width=\linewidth]{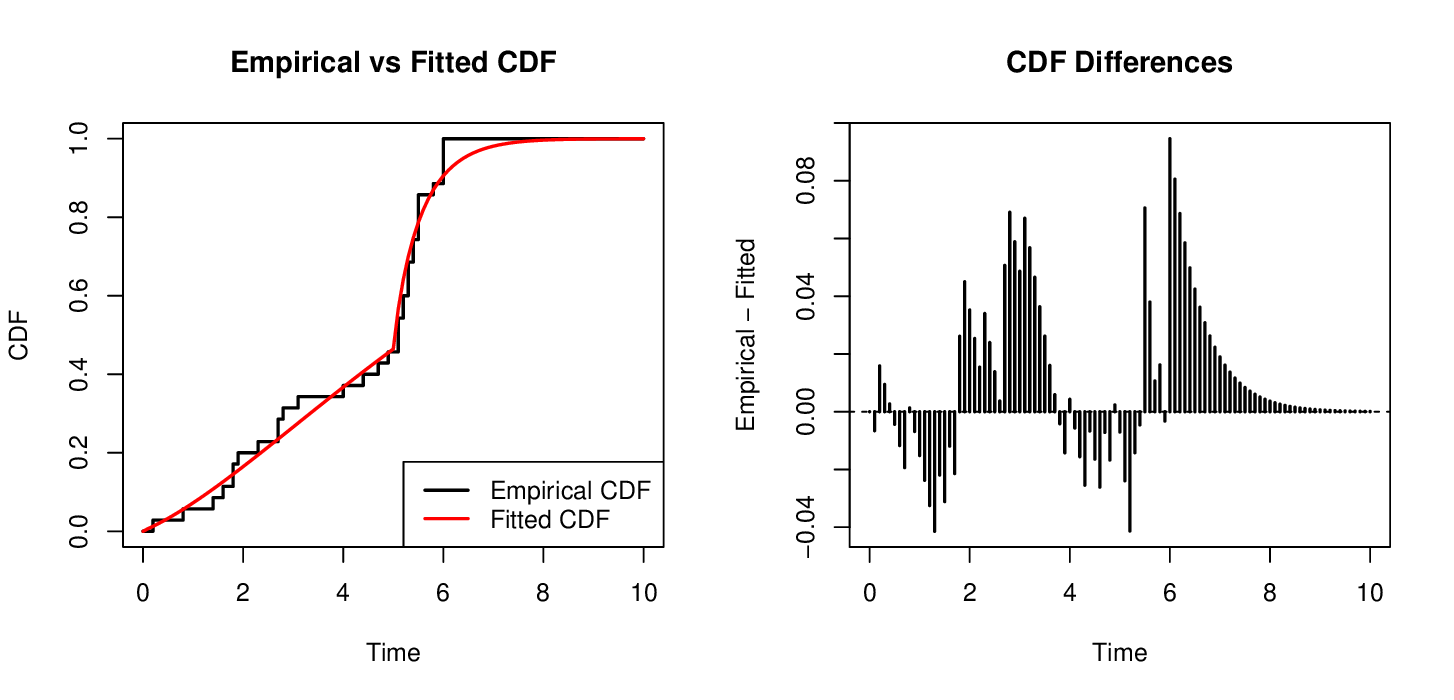}
		\caption{EDF-based goodness-of-fit diagnostics: empirical and
			model-based fitted CDFs (left panel) and their pointwise CDF
			differences (right panel).}
		\label{fig:gof_edf}
	\end{figure}

	The Kolmogorov-Smirnov~(KS) statistic was $0.0946$ with a bootstrap $p$-value 
	of $0.801$, and the Cram\'{e}r-von Mises~(CvM) statistic was $0.0372$ with a 
	bootstrap $p$-value of $0.951$. Neither test provides evidence of lack of fit, 
	indicating that the Weibull competing risks model gives an adequate description 
	of the observed failure time data. The EDF-based diagnostics are shown in 
	Figure~\ref{fig:gof_edf}.
	
	\subsection{Prior Elicitation}
	\label{subsec:prior}
	
	Specifying meaningful priors for the original model parameters 
	$(a_j, b_j, \beta_j)$ is notoriously difficult in practice, since 
	these quantities lack direct physical interpretation and are strongly 
	correlated. The reparametrisation to $\boldsymbol{\varphi}_j = 
	(t^q_{0,j}, -b_j, \beta_j)$ introduced in Section~\ref{subsec:prior_selection} 
	resolves this difficulty. The quantity $t^q_{0,j}$ is the $q$-th 
	quantile of the failure time distribution for risk $j$ at the normal 
	use stress $s_0$, which engineers can assess from field experience 
	with the device \citep{zhang2006bayesian}. The parameter $-b_j > 0$ 
	quantifies the sensitivity of lifetime to stress and can be judged 
	from knowledge of the acceleration behaviour, for instance via the 
	Arrhenius relationship \citep{meeker1998statistical}. The shape 
	parameter $\beta_j$ governs the hazard rate profile and is 
	assessable from the failure mode characteristics of each competing 
	risk, independently of the scale \citep{singpurwalla1988interactive}. Crucially, 
	the three components are approximately mutually independent, so 
	engineers can reason about them separately rather than jointly 
	\citep{zhang2006bayesian}. We exploit this structure to elicit priors from 
	the preliminary dataset of \cite{han2014inference}.
	
\begin{figure}[t]
		\centering
		\includegraphics[width=1\linewidth]{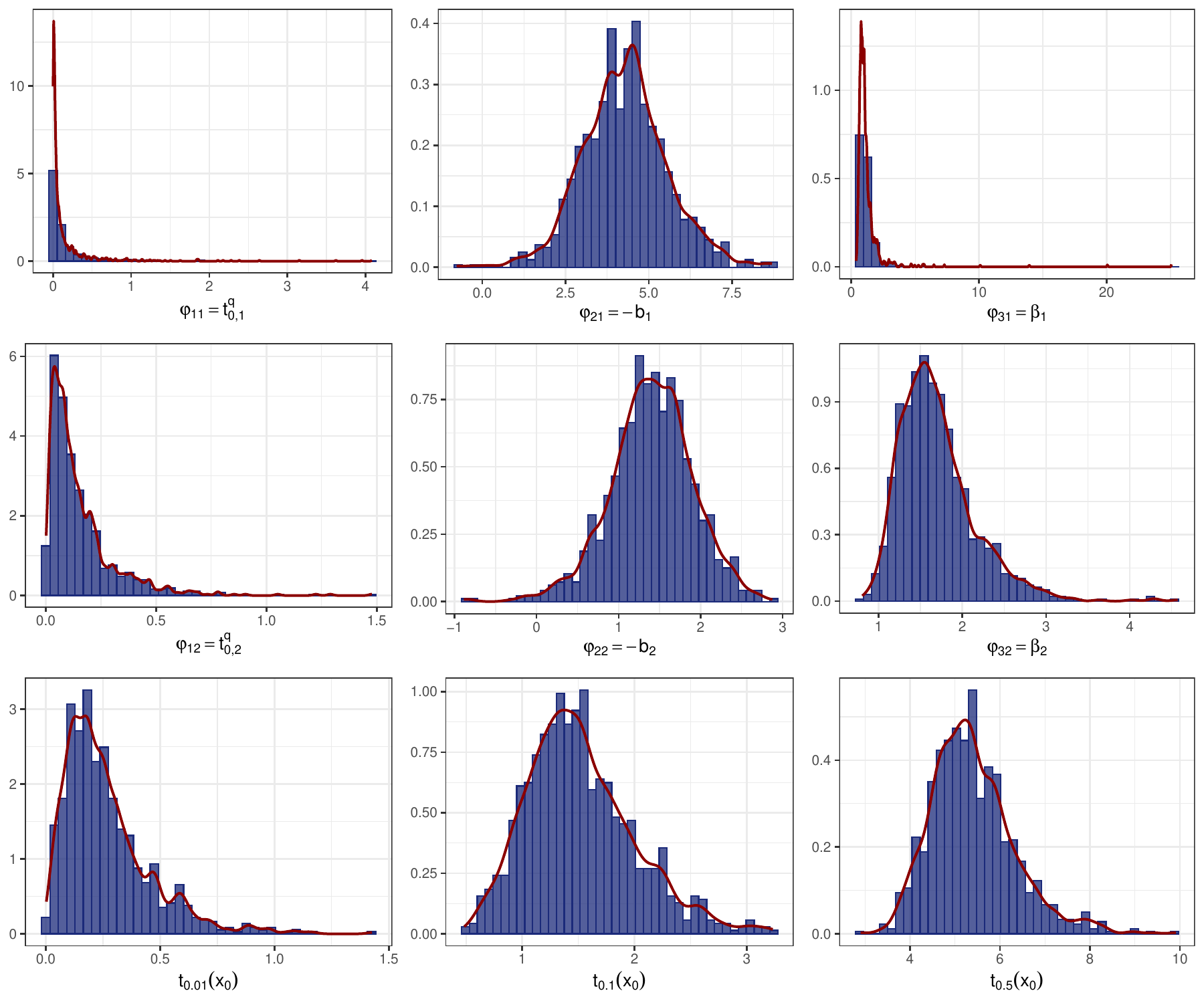}
		\caption{Histograms of the MLEs for the 
			reparametrised parameters $\boldsymbol{\varphi}_j 
			= (t^q_{0,j}, -b_j, \beta_j)$ and selected 
			quantiles $t_p(x_0)$ at use stress $s_0$, based 
			on 1000 parametric bootstrap replications.}
		\label{fig:bootstrap_hist}
	\end{figure}

	A parametric bootstrap with 1000 replicates is used to characterise 
	the sampling variability of the MLEs of $\boldsymbol{\varphi}_j$ 
	under the fitted step-stress competing risks model. For each 
	replicate, the model is refitted and the MLE of each $\varphi_{ij}$ 
	is recorded. Bootstrap replicates in which one or more cause-stress 
	combinations contained no observed failures were discarded to ensure 
	numerical stability of the likelihood, as the likelihood 
	does not have a finite maximum in such cases; 1000 valid replicates were retained in total. The 
	resulting bootstrap means, standard errors and corresponding histograms are reported in 
	Table \ref{table:bootstrap} and Figure \ref{fig:bootstrap_hist}, respectively.
	
	\begin{table}[H]
		\centering
		\caption{Bootstrap means and standard errors for the 
			reparametrised parameters $\boldsymbol{\varphi}_j 
			= (t^q_{0,j}, -b_j, \beta_j)$ and selected 
			quantiles $t_p(x_0)$ at use stress $s_0$, based 
			on 1000 parametric bootstrap replications.}
		\label{table:bootstrap}
		\begin{tabular}{lcc}
			\toprule
			Parameter & Mean & SE \\
			\midrule
			$\varphi_{11} = t^q_{0,1}$ & 0.1634 & 0.3705 \\
			$\varphi_{21} = -b_1$      & 4.2805 & 1.2737 \\
			$\varphi_{31} = \beta_1$   & 1.2006 & 1.2724 \\
			$\varphi_{12} = t^q_{0,2}$ & 0.1527 & 0.1550 \\
			$\varphi_{22} = -b_2$      & 1.4025 & 0.5039 \\
			$\varphi_{32} = \beta_2$   & 1.6989 & 0.4604 \\
			\midrule
			$t_{0.01}(x_0)$ & 0.2662 & 0.1914 \\
			$t_{0.10}(x_0)$ & 1.5195 & 0.4804 \\
			$t_{0.50}(x_0)$ & 5.3893 & 0.9250 \\
			\bottomrule
		\end{tabular}
	\end{table}
	
	Each $\varphi_{ij}$ is assigned an independent Gamma prior with 
	shape and rate parameters determined by matching the prior mean 
	and variance to the bootstrap mean $\mu_{ij}$ and standard error 
	$\sigma_{ij}$ via the method-of-moments identities
	\begin{equation}
		\alpha_{ij} = \frac{\mu_{ij}^2}{\sigma_{ij}^2}, \qquad
		\lambda_{ij} = \frac{\mu_{ij}}{\sigma_{ij}^2}.
		\label{eq:mom}
	\end{equation}
	Three priors are considered, each built using~\eqref{eq:mom} 
	but differing in the degree of confidence placed on the 
	preliminary estimates. Their hyperparameters are presented in 
	Table~\ref{table:priors}.
	
	\noindent \underline{Prior I}: Applying~\eqref{eq:mom} directly to the bootstrap summaries in 
	Table~\ref{table:bootstrap} yields Prior I. This prior is tightly 
	centred on the preliminary MLE-based estimates, with spread 
	calibrated to the actual sampling uncertainty observed in the 
	bootstrap. It represents the situation in which an engineer has 
	reasonable confidence in the preliminary dataset and is willing 
	to let it strongly inform the planned experiment. Prior I serves 
	as the baseline throughout the remainder of the paper. We refer 
	to Prior I as the tighter prior in this study.
	
	\noindent \underline{Prior II}:
	An engineer may be less confident in the precision of the 
	preliminary estimates than the bootstrap standard errors suggest, for instance, if the preliminary dataset is small, or if 
	the engineer wishes to guard against over-reliance on a single 
	historical source. Prior II accommodates this by inflating each 
	bootstrap standard error by a factor of $1.5$ before 
	applying~\eqref{eq:mom}, while leaving the prior means unchanged. 
	The factor $1.5$ roughly doubles the prior variance and produces 
	a noticeably wider distribution without making it uninformative. We refer to Prior II as the wider prior.
	The effect of prior spread on the optimal design is examined 
	in Section~\ref{subsec:sa3}.
	
	\noindent \underline{Prior III}: 
	The acceleration parameters $\varphi_{2j} = -b_j$ govern how 
	strongly lifetime responds to changes in stress and play a 
	central role in extrapolating results to use conditions. An 
	engineer may believe the true acceleration effect is stronger 
	than the preliminary data indicate. Prior III retains the wider 
	spread of Prior II but additionally shifts the prior means of 
	$\varphi_{21}$ and $\varphi_{22}$ upward by $1.5$ times their 
	respective bootstrap standard errors. Concretely, the prior mean of $\varphi_{2j}$ is shifted to 
	$\mu_{2j} + 1.5\,\sigma_{2j}$ while the spread remains 
	$1.5\,\sigma_{2j}$, with all other components identical to 
	Prior II. The effect of this shift on the optimal design is 
	examined in Section~\ref{subsec:sa3}.
	
	%	\begin{remark}
		%		\label{rem:nodoubleuse}
		%		Using a preliminary dataset for prior elicitation is standard 
		%		Bayesian practice \citep{gelman2013bayesian}. The \cite{han2014inference} 
		%		dataset informs the prior. The informative prior is 
		%		a strength of the approach rather than a limitation, it 
		%		incorporates genuine engineering knowledge and is particularly 
		%		valuable when the planned experiment involves a small sample. The 
		%		\cite{han2014inference} dataset serves solely to elicit the 
		%		prior distributions and is independent of the future experiment 
		%		whose design is being optimised.
		%	\end{remark}
	%	
	%	\begin{remark}
		%		\label{rem:qchoice}
		%		The value $q = 0.001$ is chosen small enough that $t^q_{0,j}$ 
		%		is approximately independent of the shape parameter $\beta_j$ 
		%		under heavy censoring \citep{meeker1998statistical}, while 
		%		remaining numerically stable for the parameter values 
		%		encountered in this study.
		%	\end{remark}
	
	\begin{table}[H]
		\centering
		\caption{Gamma prior hyperparameters $(\alpha_{ij}, \lambda_{ij})$ 
			under the three prior specifications considered in this 
			study. Prior I (tighter) is calibrated directly to the bootstrap 
			standard errors. Prior II (wider) uses inflated standard errors 
			(factor $1.5$) with unchanged means. Prior III modifies Prior II by shifting the prior means of 
			the acceleration parameters $\varphi_{2j}$ upward by $1.5$ 
			times their Prior~I bootstrap standard errors.}
		\label{table:priors}
		\begin{tabular}{lcccccc}
			\toprule
			& \multicolumn{2}{c}{Prior I} 
			& \multicolumn{2}{c}{Prior II} 
			& \multicolumn{2}{c}{Prior III} \\
			\cmidrule(lr){2-3}\cmidrule(lr){4-5}\cmidrule(lr){6-7}
			Parameter 
			& $\alpha_{ij}$ & $\lambda_{ij}$ 
			& $\alpha_{ij}$ & $\lambda_{ij}$
			& $\alpha_{ij}$ & $\lambda_{ij}$ \\
			\midrule
			$\varphi_{11}$ 
			& 0.195  & 1.192 & 0.086 & 0.530 & 0.086 & 0.530 \\
			$\varphi_{21}$ & 11.290 & 2.637 & 5.018 & 1.172 & 10.501 & 1.696 \\
			$\varphi_{31}$ 
			& 0.889  & 0.741 & 0.395 & 0.329 & 0.395 & 0.329 \\
			$\varphi_{12}$ 
			& 0.970  & 6.354 & 0.431 & 2.824 & 0.431 & 2.824 \\
			$\varphi_{22}$ & 7.748  & 5.526 & 3.444 & 2.456 & 8.153  & 3.778 \\
			$\varphi_{32}$ 
			& 13.606 & 8.012 & 6.047 & 3.561 & 6.047 & 3.561 \\
			\bottomrule
		\end{tabular}
	\end{table}
	
	Prior I serves as the reference specification throughout the 
	numerical study. Convergence of the NUTS sampler under Prior I 
	is examined in Section~\ref{sec:conv_diag} before the sensitivity 
	results are presented, since reliable posterior estimates are a 
	prerequisite for meaningful criterion evaluation. The sensitivity 
	of the optimal design to the choice of prior is investigated in 
	Section~\ref{subsec:sa3}, where results under all three priors 
	are compared.
	
	\begin{table*}[tbp]
		\centering
		\scriptsize
		\begin{tabular}{@{}lrrrrrrrrr@{}}
			\toprule
			variable & mean & median & sd & mad & q5 & q95 & rhat & ess\_bulk & ess\_tail \\ 
			\midrule
			lp\_\_ & -94.75 & -94.38 & 1.78 & 1.59 & -98.14 & -92.52 & 1.00 & 1256 & 1683 \\
			
			$\varphi_{11}$ & 0.46 & 0.35 & 0.40 & 0.31 & 0.04 & 1.29 & 1.00 & 1515 & 1366 \\
			$\varphi_{21}$ & 4.71 & 4.68 & 1.08 & 1.07 & 2.96 & 6.56 & 1.00 & 1809 & 1493 \\
			$\varphi_{31}$ & 1.28 & 1.25 & 0.29 & 0.28 & 0.86 & 1.79 & 1.00 & 1339 & 1379 \\		
			$\varphi_{12}$ & 0.12 & 0.11 & 0.07 & 0.06 & 0.04 & 0.24 & 1.00 & 1525 & 1623 \\
			$\varphi_{22}$ & 1.39 & 1.33 & 0.46 & 0.43 & 0.73 & 2.20 & 1.00 & 2166 & 1931 \\
			$\varphi_{32}$ & 1.62 & 1.61 & 0.21 & 0.20 & 1.29 & 1.98 & 1.00 & 1528 & 1719 \\		
			
			$a_1$ & 4.52 & 4.45 & 0.77 & 0.75 & 3.37 & 5.83 & 1.00 & 1719 & 1709 \\
			$b_1$ & -4.71 & -4.68 & 1.08 & 1.07 & -6.56 & -2.96 & 1.00 & 1809 & 1493 \\
			$\beta_1$ & 1.28 & 1.25 & 0.29 & 0.28 & 0.86 & 1.79 & 1.00 & 1339 & 1379 \\
			$a_2$ & 2.06 & 2.04 & 0.27 & 0.26 & 1.65 & 2.57 & 1.00 & 2156 & 2056 \\
			$b_2$ & -1.39 & -1.33 & 0.46 & 0.43 & -2.20 & -0.73 & 1.00 & 2166 & 1931 \\
			$\beta_2$ & 1.62 & 1.61 & 0.21 & 0.20 & 1.29 & 1.98 & 1.00 & 1528 & 1719 \\
			
			$\log(\theta_1{(x_1)})$ & 2.16 & 2.12 & 0.35 & 0.32 & 1.70 & 2.78 & 1.00 & 2108 & 1848 \\
			$\log(\theta_1{(x_2)})$ & -0.19 & -0.20 & 0.49 & 0.48 & -0.95 & 0.60 & 1.00 & 2195 & 2028 \\
			$\log(\theta_2{(x_1)})$ & 1.36 & 1.36 & 0.13 & 0.13 & 1.16 & 1.59 & 1.00 & 2929 & 1908 \\
			$\log(\theta_2{(x_2)})$ & 0.67 & 0.69 & 0.26 & 0.24 & 0.22 & 1.06 & 1.00 & 2525 & 2326 \\
			
			$\theta_1{(x_1)}$ & 9.32 & 8.30 & 4.35 & 2.58 & 5.47 & 16.18 & 1.00 & 2108 & 1848 \\
			$\theta_1{(x_2)}$ & 0.93 & 0.82 & 0.53 & 0.39 & 0.39 & 1.82 & 1.00 & 2195 & 2028 \\
			$\theta_2{(x_1)}$ & 3.94 & 3.89 & 0.53 & 0.50 & 3.18 & 4.90 & 1.00 & 2929 & 1908 \\
			$\theta_2{(x_2)}$ & 2.02 & 1.99 & 0.50 & 0.49 & 1.24 & 2.90 & 1.00 & 2525 & 2326 \\
			
			$t_p(x_0)$ & 1.84 & 1.77 & 0.54 & 0.46 & 1.11 & 2.82 & 1.00 & 1773 & 1648 \\
			$\log(t_p(x_0))$ & 0.57 & 0.57 & 0.28 & 0.26 & 0.11 & 1.04 & 1.00 & 1773 & 1648 \\
			
			\bottomrule
		\end{tabular}
		\caption{Convergence diagnosis: Posterior summary statistics of model parameters.}
		\label{table2}
	\end{table*}
	
	\section{Convergence Diagnosis}
	\label{sec:conv_diag}
	
	To present a diagnosis on convergence of the estimates, $\tau$ and $s_1$ are fixed to $5$ (in 100 hours) $(0.8333)$ and $1/320.2136~\text{K}^{-1}$ $(0.5)$, respectively. The quantities in brackets denote the corresponding standardised values. We take $3$ chains where 
	each chain contains $2000$ iterations, discarding the first $1000$ 
	as warm-up. Initial values for each chain are drawn independently at random 
	from the Gamma prior distributions of the $\varphi_{ij}$ parameters, 
	ensuring that the chains start from dispersed positions in the 
	parameter space and providing a natural check on convergence: if 
	all chains reach the same stationary distribution from diverse 
	initialisations, this constitutes strong evidence that the sampler 
	has converged to the correct posterior. The overall seed number and the seed number inside 
	\texttt{mod\$sample()} are set to $2026$ and $135$, respectively, 
	and no thinning is applied in the \texttt{cmdstanr} package of R. 
	Increasing the number of chains and iterations will no doubt 
	enhance the performance further.
	
\begin{figure}[tbp]
		\centering
		\begin{subfigure}{1\linewidth}
			\includegraphics[width=1\linewidth, height=0.65\linewidth]{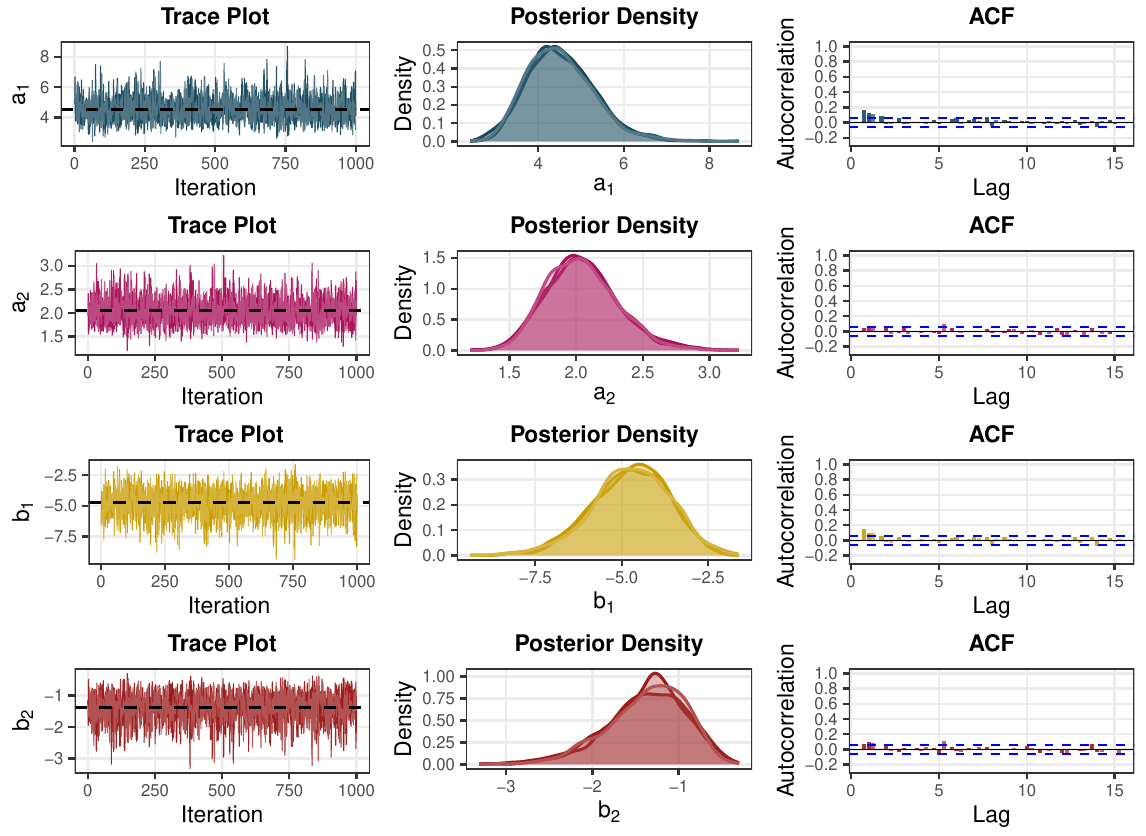}
			\label{fig:sub1}
		\end{subfigure} 
		\begin{subfigure}{1\linewidth}
			\includegraphics[width=1\linewidth, height=0.65\linewidth]{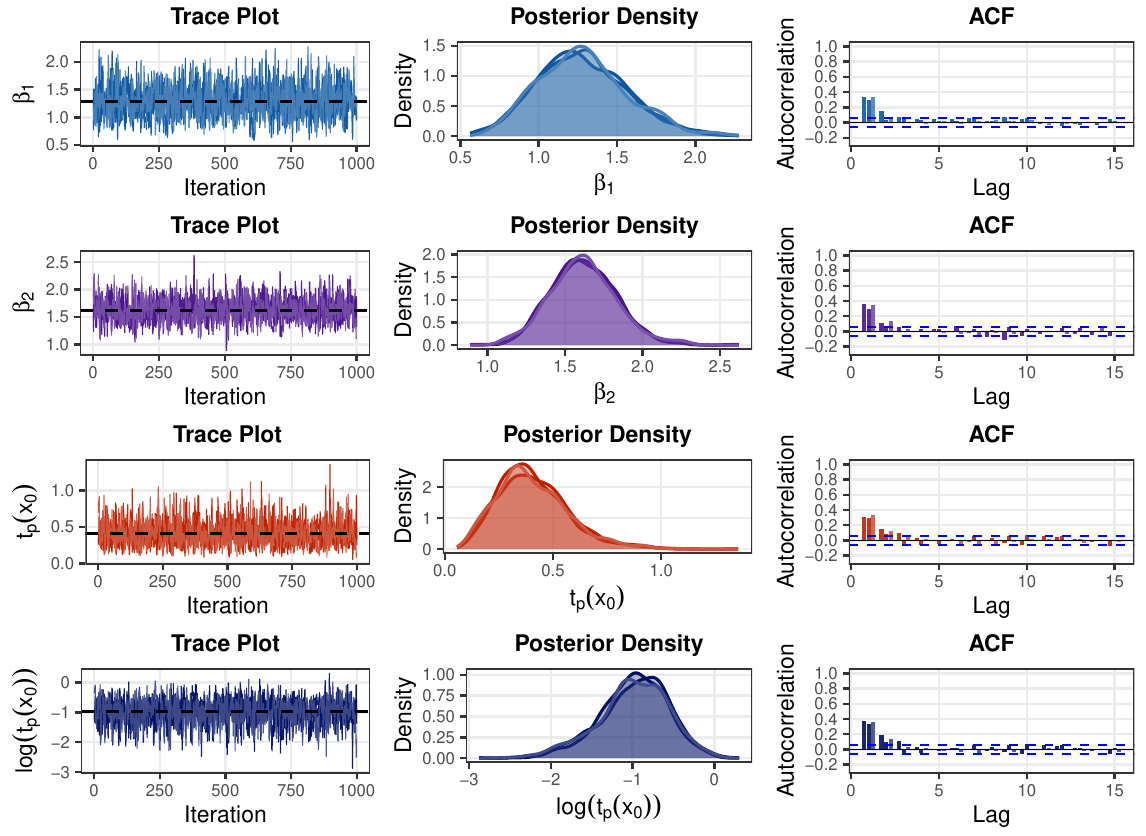}
			\label{fig:sub2}
		\end{subfigure}
		\caption{Convergence diagnosis: Trace, density, and ACF plots.}
		\label{fig2}
	\end{figure}

	Table~\ref{table2} reports the posterior summary statistics for all 
	model parameters, including the posterior mean, median, standard 
	deviation, median absolute deviation, 5th and 95th percentiles, 
	$\hat{R}$, and the bulk and tail effective sample sizes. The 
	$\hat{R}$ values are equal to $1.00$ for all parameters, and we 
	also verified that the number of divergent transitions and the 
	number of iterations exceeding the maximum treedepth are both $0$, 
	together confirming convergence of all three chains. In addition, 
	\texttt{ess\_bulk} and \texttt{ess\_tail} lie in the range 
	$[1256,\, 2929]$ and $[1366,\, 2326]$, respectively, across all 
	parameters, indicating that the effective sample sizes are adequate 
	for reliable posterior inference. The trace, posterior density plots, autocorrelations over lags of the 
	estimates are demonstrated for each chain and reported in 
	Figure~\ref{fig2}. The trace plots exhibit good mixing across all three chains. The posterior 
	density plots are smooth and approximately unimodal for all 
	parameters, suggesting that the chains have adequately explored 
	the bulk of the posterior. The autocorrelation function (ACF) plots 
	show rapid decay to near zero within the first few lags for all 
	parameters, confirming that successive draws within each chain are 
	nearly independent. Therefore, the reported results and plots are satisfactory 
	for confirming the convergence of the obtained posterior estimates.

	%			\begin{figure}[tbp]
		%				\centering
		%				\begin{subfigure}{0.85\linewidth}
			%					\includegraphics[width=1\linewidth, height=0.65\linewidth]{posterior_density_plots_prior_I_1.eps}
			%					\label{fig:sub1}
			%				\end{subfigure} 
		%				\begin{subfigure}{0.85\linewidth}
			%					\includegraphics[width=1\linewidth, height=0.65\linewidth]{posterior_density_plots_prior_I_2.eps}
			%					\label{fig:sub2}
			%				\end{subfigure}
		%				\caption{Case study: Posterior plots}
		%				\label{fig2}
		%			\end{figure}
	%			
	%			%\begin{landscape}
	%			\begin{figure}[tbp]
		%				\centering
		%				\includegraphics[width=1\linewidth]{acf_prior_I.pdf}
		%				\caption{Case study: Autocorrelation plots}
		%				\label{fig3}
		%			\end{figure}
	
	%\end{landscape}
	%\begin{figure}[tbp]
	%	\centering
	%	\begin{subfigure}[b]{0.5\linewidth}
		%		\centering
		%		\includegraphics[width=\linewidth]{MA_tp_135_priorI.eps}
		%		\caption{$t_p(x_0)$}
		%	\end{subfigure}%
	%	\begin{subfigure}[b]{0.5\linewidth}
		%		\centering
		%		\includegraphics[width=\linewidth]{MA_log_tp_135_priorI.eps}
		%		\caption{$\log(t_p(x_0))$}
		%	\end{subfigure}
	%	\caption{Case study: Plots of moving averages}
	%	\label{fig4}
	%\end{figure}

	% ============================================================
	% Section: Sensitivity Analysis
	% One-variable optimal design (optimising over tau)
	% Four sub-analyses: p, s1, prior, n
	% ============================================================
	
	It is worth mentioning that convergence issues can sometimes arise 
	during the Monte Carlo loop used for computing the criterion 
	functions. For instance, a few posterior draws may behave like 
	outliers within a chain, or the chains may fail to initialise 
	properly due to unfavourable starting values drawn from the gamma 
	priors. To safeguard against such issues, our algorithm implements 
	an automated refit strategy: after each initial fit, the 
	diagnostics are checked for divergent transitions, $\hat{R} > 1.01$, 
	infinite or excessively large posterior standard deviations, and 
	insufficient effective sample sizes in both bulk and tail. If any 
	of these conditions are met, the fit is automatically discarded and 
	rerun with more conservative sampler settings, specifically 
	\texttt{iter\_warmup} $= 2000$, \texttt{iter\_sampling} $= 2000$, 
	\texttt{adapt\_delta} $= 0.99$, and \texttt{max\_treedepth} $= 15$. 
	If the refit still fails to meet all diagnostic criteria, the 
	Monte Carlo replicate is discarded entirely and not used in the 
	criterion evaluation. %Figure~\ref{fig6} shows the surface plots of $C_1(D)$ and $C_2(D)$ before applying this correction, illustrating the effect of unconverged fits on the criterion surface.
	
	\section{Sensitivity Analysis}
	\label{sec:sa}
	\subsection{One-Variable Optimal Design}
	In this section we investigate the robustness of the 
	one-variable optimal design, the optimal stress-change 
	time $\tau^0$ obtained by minimising $C_1(D)$ and 
	$C_2(D)$ over $\tau$ for a fixed lower stress level 
	$s_1$, with respect to four sources of variation: the 
	quantile probability $p$, the lower stress level $s_1$, 
	the prior hyperparameter specification, and the sample 
	size $n$.
	
	In each sub-analysis, exactly one quantity is varied 
	while all remaining parameters are held fixed at their 
	baseline values. The baseline configuration is as 
	follows: quantile probability $p = 0.10$, lower stress 
	level $s_1 = 1/320.2136$~K$^{-1}$ (standardised 
	$x_1 = 0.50$, the midpoint of the feasible stress 
	range), Prior I as elicited in 
	Section~\ref{subsec:prior}, sample size $n = 35$, 
	experimental duration $t_c = 6$ (hundred hours), highest 
	stress $s_2 = 1/353$~K$^{-1}$, use stress 
	$s_0 = 1/293$~K$^{-1}$, and $B = 1000$ Monte Carlo 
	replications per grid point. The $\tau$ grid consists 
	of $M_1 = 25$ equally spaced points over the interval 
	$(0.050, 5.950)$. The lower stress grid used in SA2 (\ref{subsec:sa2}) is constructed as the 
	nine interior points of an equally spaced sequence from 
	$s_0 = 1/293$~K$^{-1}$ to $s_2 = 1/353$~K$^{-1}$ such 
	that $x_1=0.1(0.1)0.9$, \textit{i.e.},
	\begin{align}
		s_1^{-1} \in  \{ & 298.0663,~303.3109,~308.7433,~
		314.3739,~320.2136,~326.2744,~332.5691,~339.1115,~
		345.9164\}~\text{K}.&
		\label{eq:s1grid}
	\end{align}
	The midpoint $320.2136$~K corresponds exactly to 
	$x_1 = 0.50$ and serves as the common baseline for 
	SA1, SA3, and SA4. Throughout the remainder of this paper, the lower stress 
	values appearing in the grid of~\eqref{eq:s1grid} are 
	displayed in their rounded forms (e.g., $320$~K instead 
	of $320.2136$~K) to improve readability in tables, figures, 
	and running text. All computations, however, are carried out 
	using the precise decimal values given in~\eqref{eq:s1grid}.
	
	\subsubsection{SA1: Sensitivity with Respect to Quantile
		Probability $p$}
	\label{subsec:sa1}
	
	The quantile probability $p$ determines the reliability
	characteristic that the designed experiment is intended to
	estimate most precisely. Three values are considered:
	$p \in \{0.01,\; 0.10,\; 0.50\}$, corresponding to the
	$1$st percentile (early-life, rare failures), the $10$th
	percentile (moderate lower-tail), and the median lifetime,
	respectively. All other parameters are fixed at the baseline:
	$s_1 = 1/320$~K$^{-1}$, Prior I, $n = 35$.
	
\begin{figure}[H]
		\centering
		\includegraphics[width=0.85\linewidth, height=0.25\textheight, keepaspectratio=false]{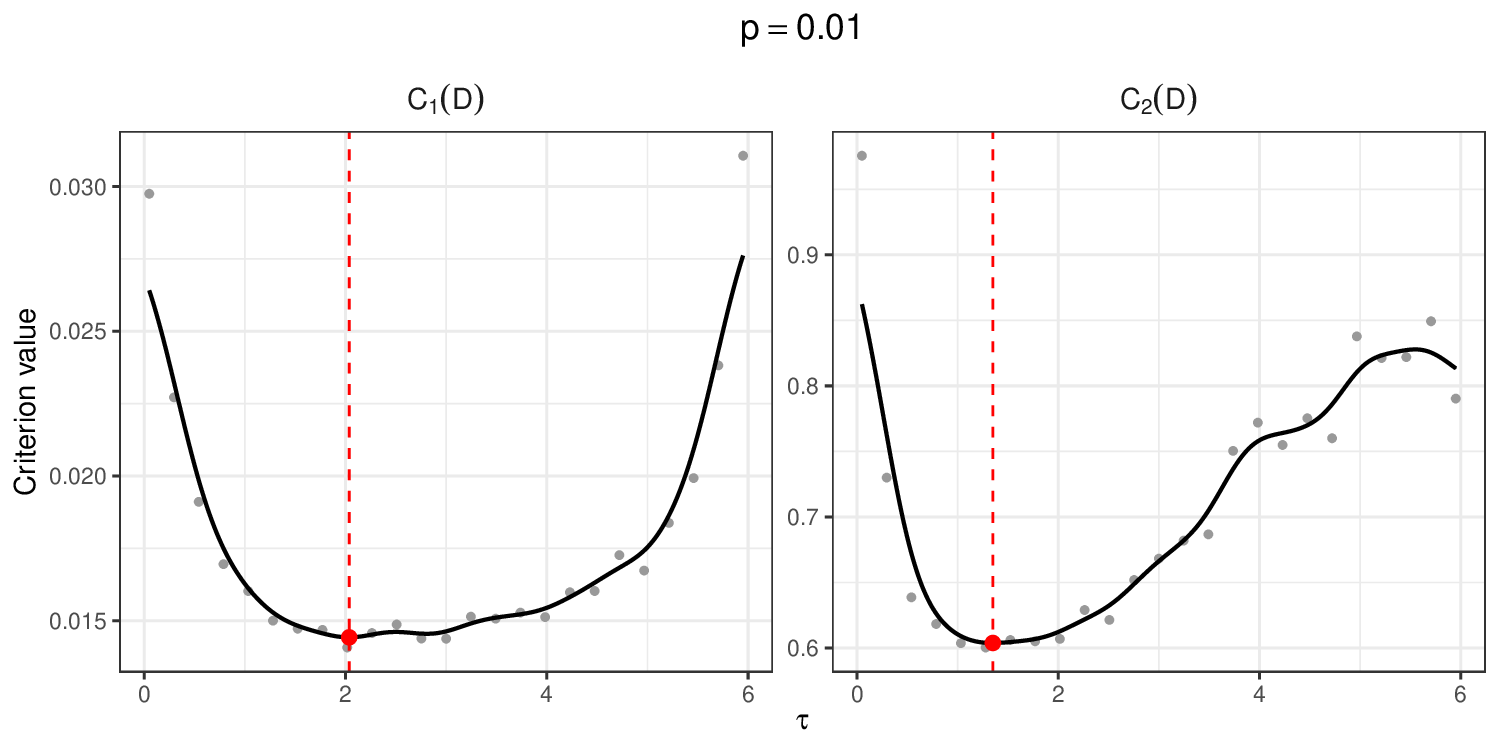}\\[4pt]
		\includegraphics[width=0.85\linewidth, height=0.25\textheight, keepaspectratio=false]{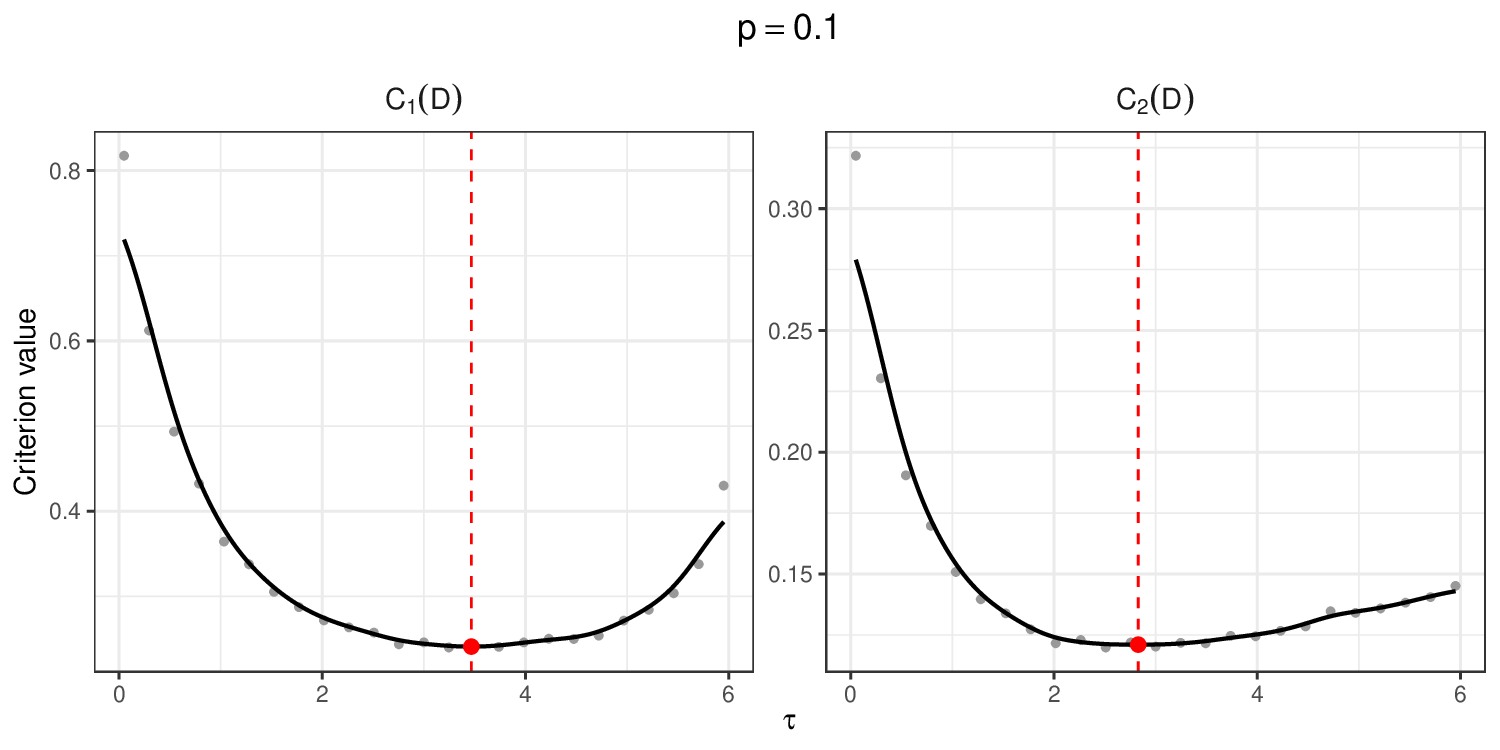}\\[4pt]
		\includegraphics[width=0.85\linewidth, height=0.25\textheight, keepaspectratio=false]{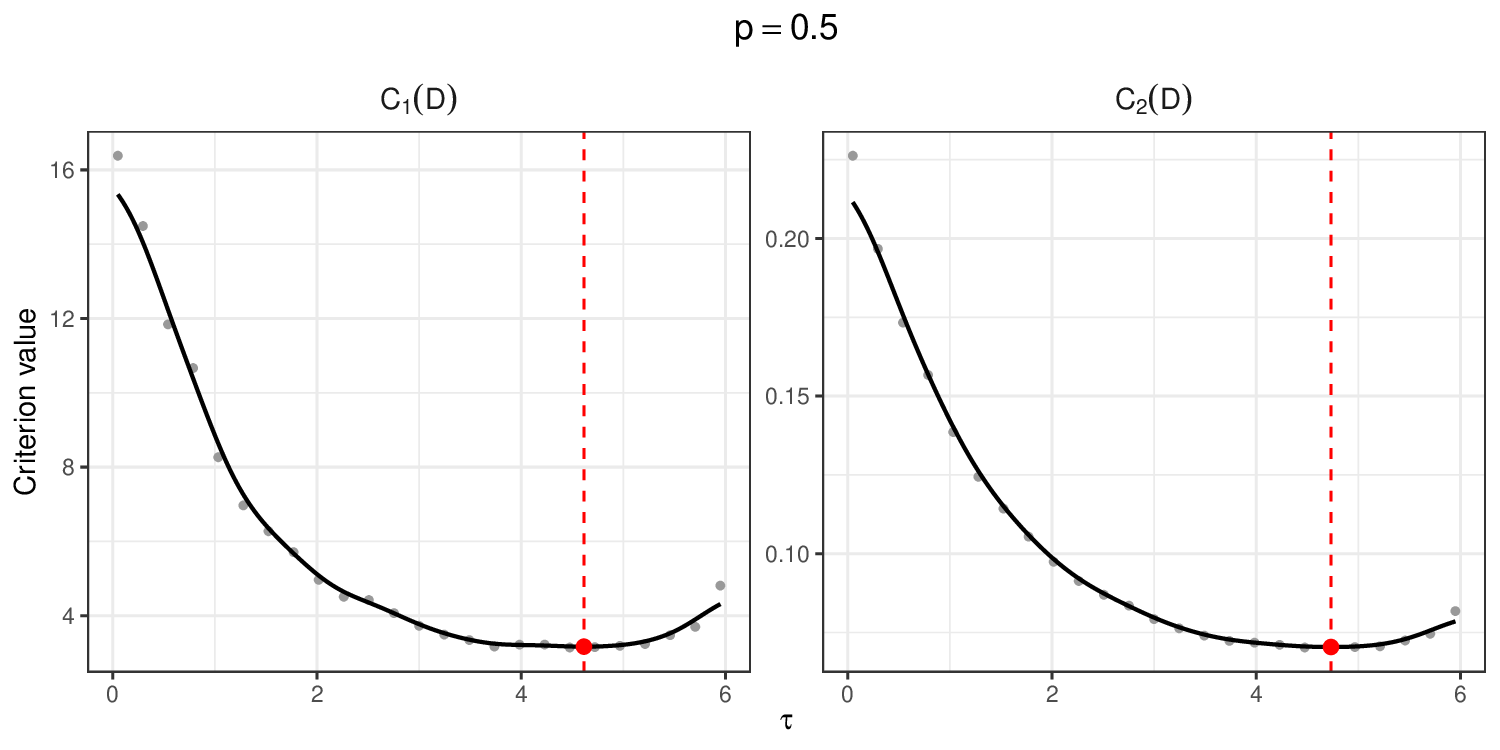}
		\caption{SA1: Smoothed criterion functions $C_1(\tau)$ and $C_2(\tau)$
			under varying quantile probability $p \in \{0.01, 0.10, 0.50\}$.
			Vertical dashed lines indicate the optimal $\tau^0$ for each criterion.}
		\label{fig:sa1}
	\end{figure}

	The estimated $p$-th quantiles of the lifetime distribution
	at use stress $s_0 = 293$~K, obtained from the bootstrap
	analysis of Section~\ref{subsec:prior}, are
	$t_{0.01}(x_0) = 0.266$, $t_{0.10}(x_0) = 1.520$, and
	$t_{0.50}(x_0) = 5.389$ (in hundred hours). These span
	a wide range, reflecting the highly skewed nature of the
	competing-risks lifetime distribution for this device.
	The bootstrap distributions of $t_p(x_0)$ for the three 
	quantile levels, shown in Figure~\ref{fig:bootstrap_hist}, 
	illustrate the considerable difference in magnitude across 
	quantile levels, further motivating the use of the 
	scale-free criterion $C_2(D)$ for comparison across 
	different values of $p$. The results on optimal designs from SA1 are reported in Table~\ref{table:sa1} and Figure~\ref{fig:sa1}. In Figure~\ref{fig:sa1}, the grey points represent the raw criterion values at each of the $M_1 = 25$ 
	discrete $\tau$ grid points, the solid black curve is 
	the Gaussian kernel smoothed criterion function, and the 
	red dashed vertical line with red dot indicates the 
	smoothed optimal $\tau^0$.
	
	\begin{table}[bp]
		\centering
		\caption{SA1: Smoothed one-variable optimal designs under varying quantile probability $p \in \{0.01, 0.10, 0.50\}$, with $s_1 = 1/320$~K$^{-1}$ ($x_1 = 0.5$), $n = 35$, Prior I.}
		\label{table:sa1}
		\begin{tabular}{ccccc}
			\toprule
			& \multicolumn{2}{c}{Criterion I} & \multicolumn{2}{c}{Criterion II} \\
			\cmidrule(lr){2-3} \cmidrule(lr){4-5}
			$p$ & $\tau^0$ & $C_1^0(\tau^0)$ & $\tau^0$ & $C_2^0(\tau^0)$ \\
			\midrule
			0.01 & 2.036 & 0.014 & 1.351 & 0.604  \\
			0.10 & 3.467 & 0.241 & 2.829 & 0.121 \\
			0.50 & 4.614 & 3.167 & 4.732 & 0.071 \\
			\bottomrule
		\end{tabular}
	\end{table}

	The optimal stress-change time $\tau^0$ increases with $p$
	for both criteria, from around $\tau^0 \approx 2$ at
	$p = 0.01$ to $\tau^0 \approx 4.6$ at $p = 0.50$ under
	Criterion I. This pattern is physically sensible. For small
	$p$, the target quantity $t_p(x_0)$ lies deep in the lower
	tail of the lifetime distribution, a region dominated by
	early failures. Cause~1 has a decreasing hazard rate
	($\hat{\beta}_1 = 0.769 < 1$) and tends to produce early
	failures, so the most informative experiment for estimating
	$t_{0.01}(x_0)$ is one that captures these early events
	efficiently, favouring an earlier switch to the higher
	stress. For larger $p$, the target shifts toward the body
	of the distribution where later failures from cause~2
	(increasing hazard, $\hat{\beta}_2 = 1.532 > 1$) become
	more informative. Accumulating these failures requires
	keeping units at $s_1$ for longer before switching, which
	pushes $\tau^0$ upward.
	
	The behaviour of the criterion values is similarly instructive.
	Under Criterion I, $C_1^0(\tau^0)$ increases sharply with
	$p$, from $0.014$ at $p = 0.01$ to $3.167$ at $p = 0.50$.
	This reflects the scale of $t_p(x_0)$ itself: the median
	lifetime is around $5.4$ hundred hours while the $1$st
	percentile is only $0.27$ hundred hours, so the absolute
	posterior variance is naturally much larger for the median.
	Under Criterion II the pattern reverses:
	$C_2^0(\tau^0)$ decreases from $0.604$ at $p = 0.01$ to
	$0.071$ at $p = 0.50$. For $p = 0.01$, the target quantile
	falls in the extreme lower tail where very few failures are
	observed under the experimental conditions. While the
	absolute variance of such a small quantity is modest, its
	logarithm is highly sensitive to small estimation errors,
	leading to a large preposterior variance on the log scale.
	The choice of $p$ should therefore reflect what the 
	engineer actually needs to know about the device, 
	whether that is early-life failure behaviour, a moderate 
	lower-tail quantile, or the median lifetime.
	
	\subsubsection{SA2: Sensitivity with Respect to Lower Stress
		Level $s_1$}
	\label{subsec:sa2}
	
	The lower stress level $s_1$ determines the thermal 
	conditions experienced by units during the first phase 
	of the step-stress experiment. In practice, $s_1$ is a 
	design choice that must balance the need to accumulate 
	failures (favouring higher $s_1$) against the goal of 
	collecting observations close to use conditions 
	(favouring lower $s_1$). The nine stress levels in 
	\eqref{eq:s1grid} are considered, covering the full 
	feasible range from $s_0 = 1/293$~K$^{-1}$ to 
	$s_2 = 1/353$~K$^{-1}$ with equal spacing in 
	temperature, excluding the boundary points. All other 
	parameters are fixed: $p = 0.10$, Prior I, $n = 35$.
	
	%The results of SA2 are reported in 
	%Table~\ref{table:sa2_priorI} and 
	%Figure~\ref{fig:sa2_smooth}. As $s_1$ increases toward 
	%$s_2$, the stress contrast between the two phases of 
	%the experiment diminishes, \textbf{reducing the information 
		%available for estimating the acceleration parameters 
		%$b_j$. Consequently,} both $C_1^0(\tau^0)$ and 
	%$C_2^0(\tau^0)$ increase monotonically with $s_1$, 
	%as confirmed by Table~\ref{table:sa2_priorI}. For 
	%Criterion II, the optimal $\tau^0$ decreases 
	%monotonically from $5.950$ at $x_1 = 0.1$ to $1.055$ 
	%at $x_1 = 0.9$, \textbf{reflecting that as $s_1$ approaches 
		%$s_2$ the experiment benefits from switching to the 
		%higher stress phase earlier.} For Criterion I, a broadly 
	%similar decreasing trend is observed from $x_1 = 0.1$ 
	%to $x_1 = 0.7$, however the optimal $\tau^0$ rises 
	%again at $x_1 = 0.8$ and $x_1 = 0.9$. This 
	%non-monotone behaviour at high $x_1$ is attributable 
	%to the flatter criterion surface when $s_1$ is close 
	%to $s_2$, as visible in the bottom panels of 
	%Figure~\ref{fig:sa2_smooth}, where the minimum of 
	%$C_1(\tau)$ becomes less well-defined. The criterion 
	%values at $x_1 = 0.9$ are substantially larger than 
	%at lower stress levels for both criteria, confirming 
	%that operating the first phase close to the upper 
	%stress $s_2$ is informationally inefficient.
	
\begin{figure}[tbp]
		\centering
		\includegraphics[width=0.7\linewidth, height=0.185\textheight, keepaspectratio=false]{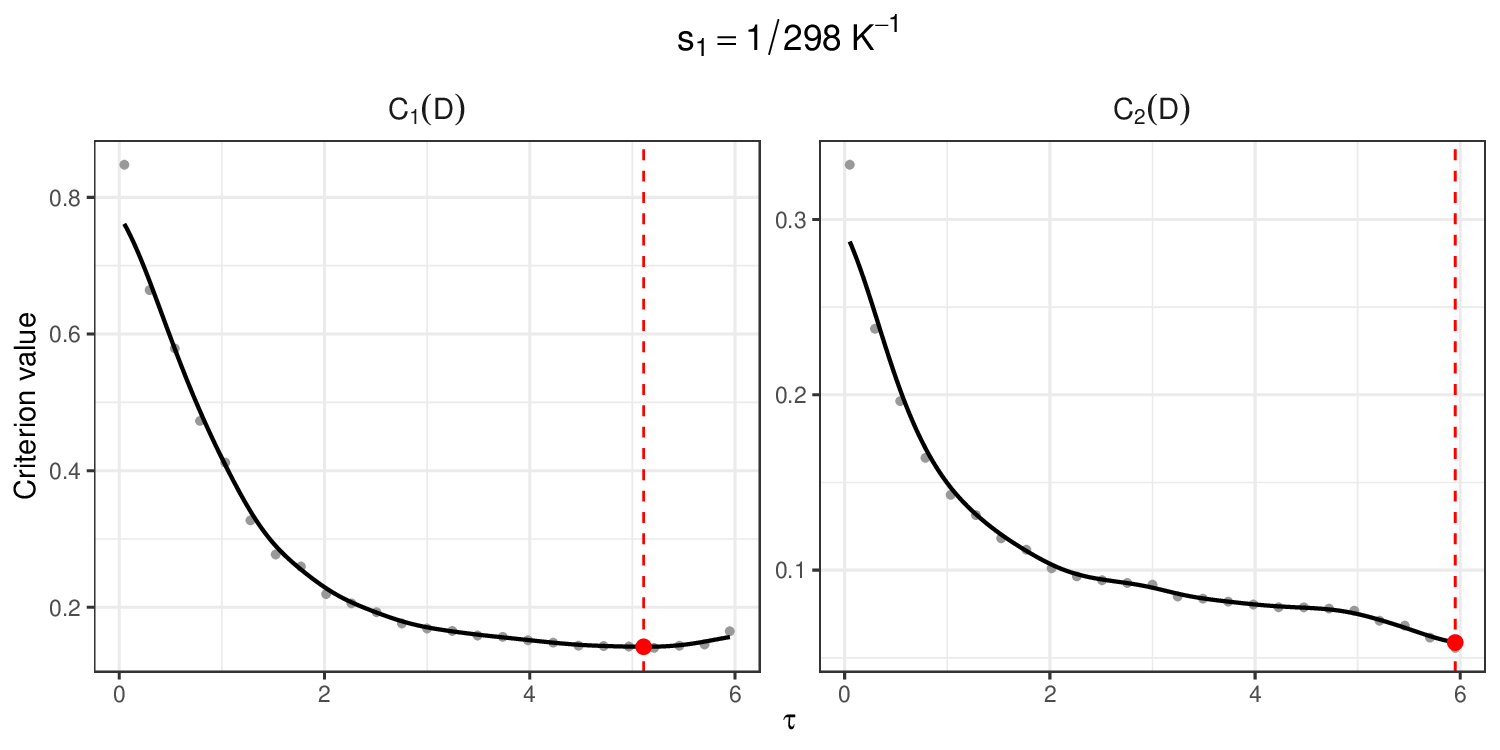}\\[2pt]
		\includegraphics[width=0.7\linewidth, height=0.185\textheight, keepaspectratio=false]{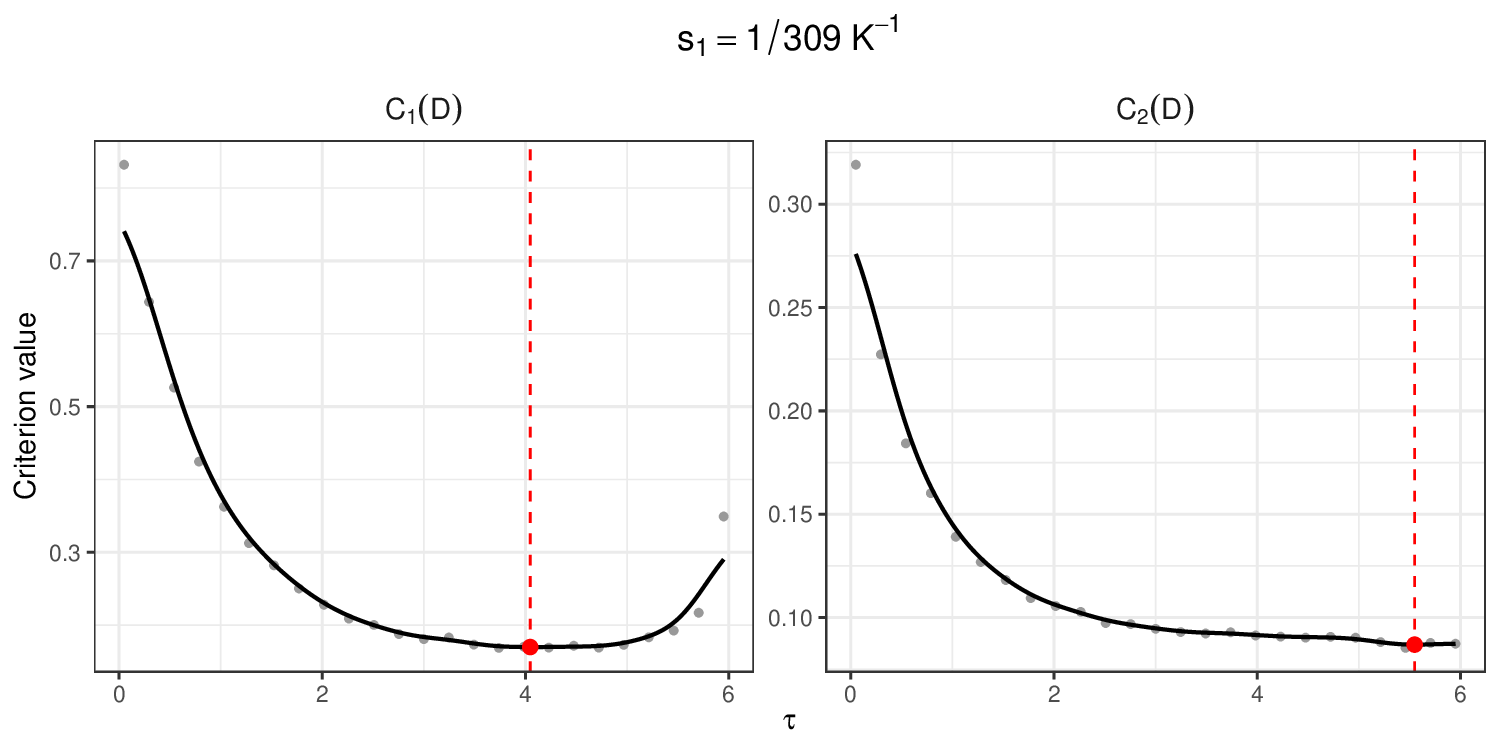}\\[2pt]
		\includegraphics[width=0.7\linewidth, height=0.185\textheight, keepaspectratio=false]{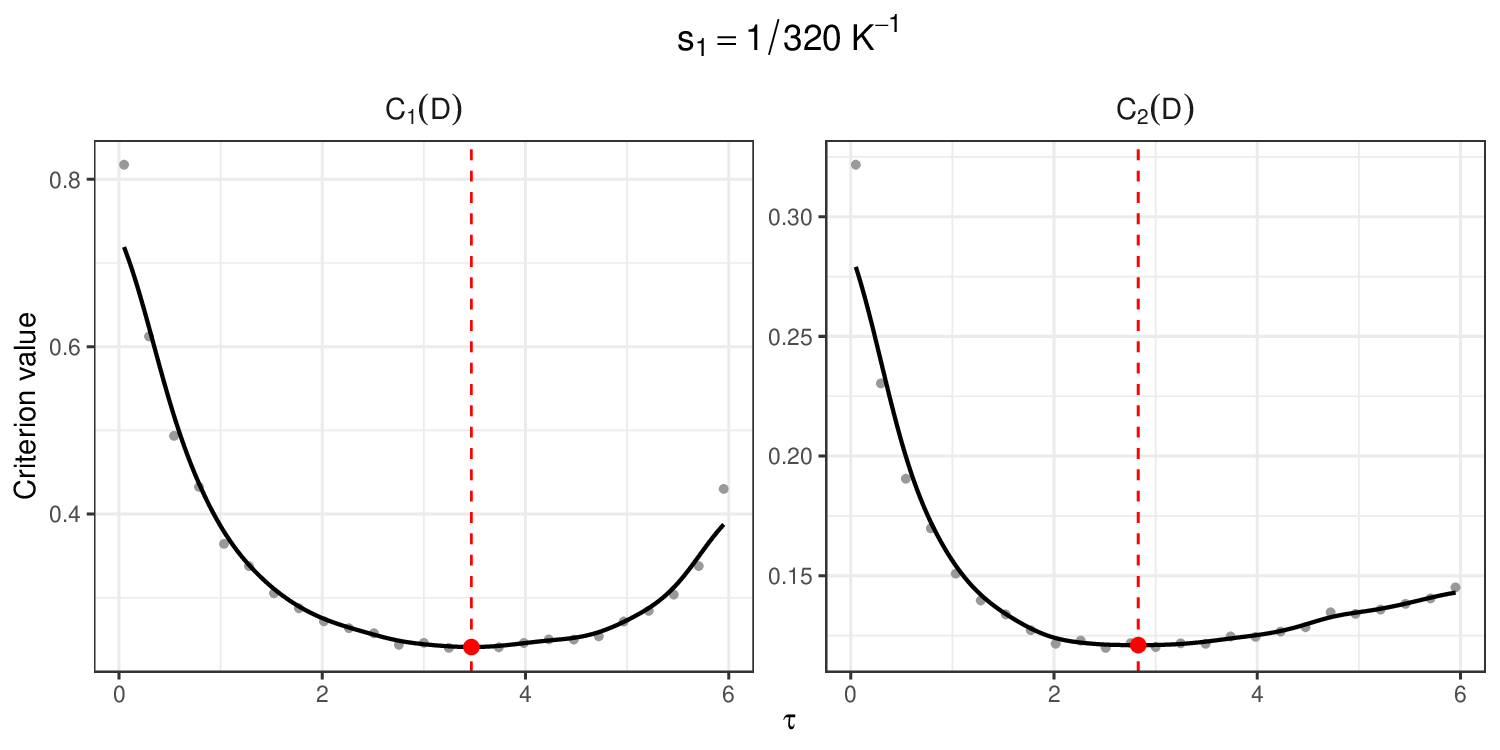}\\[2pt]
		\includegraphics[width=0.7\linewidth, height=0.185\textheight, keepaspectratio=false]{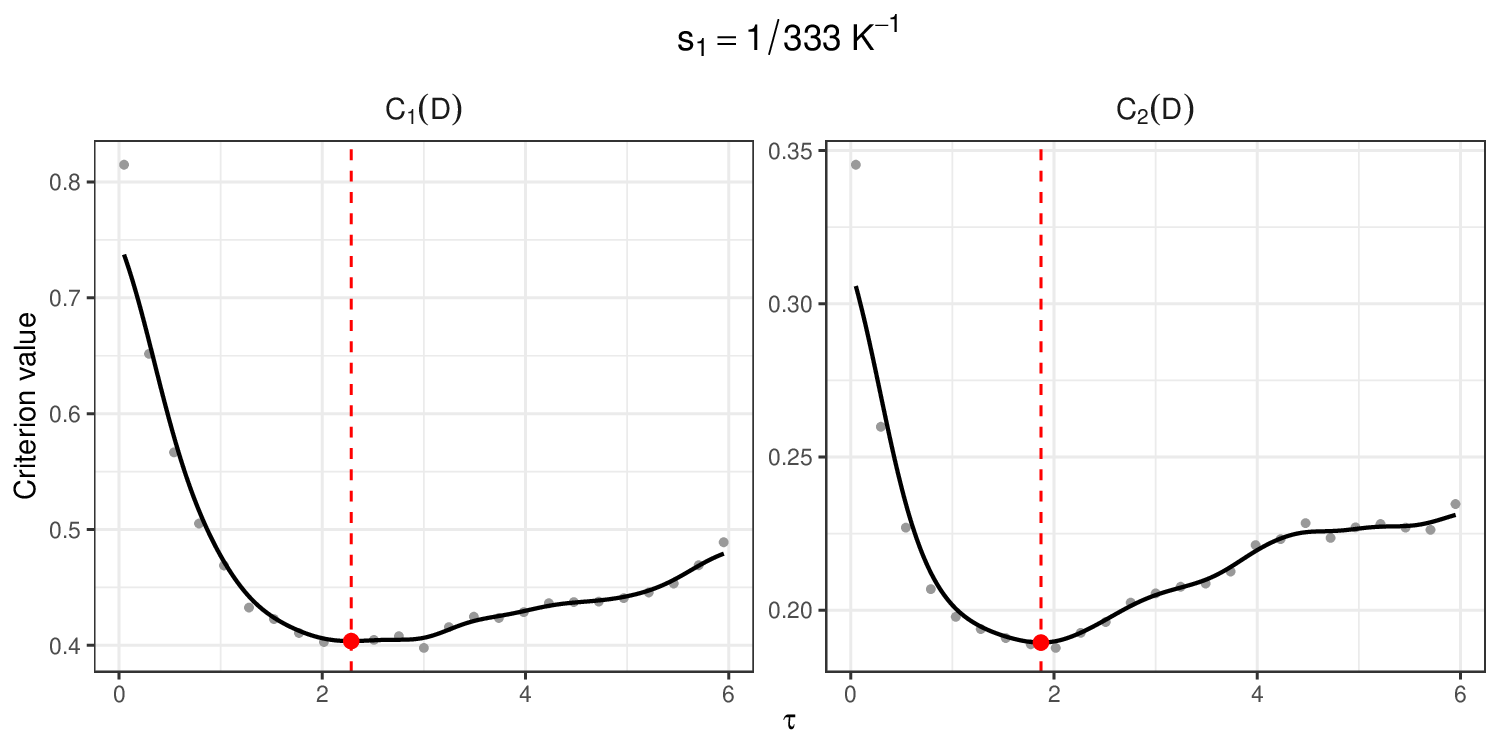}\\[2pt]
		\includegraphics[width=0.7\linewidth, height=0.185\textheight, keepaspectratio=false]{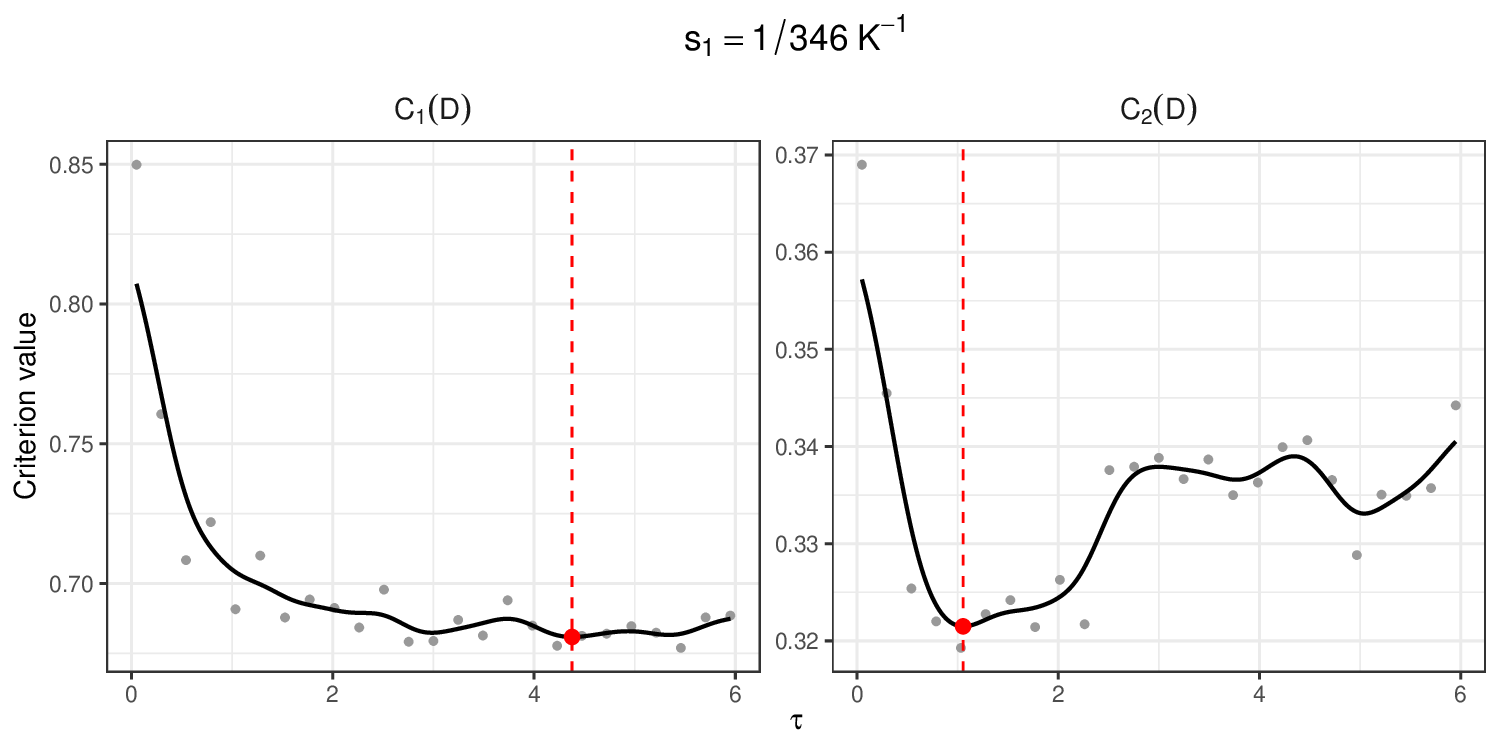}
		\caption{SA2: Smoothed criterion functions $C_1(\tau)$ and $C_2(\tau)$ 
			for five representative lower stress levels 
			$s_1 \in \{1/298, 1/309, 1/320, 1/333, 1/346\}$~K$^{-1}$. Vertical red dashed lines indicate the optimal $\tau^0$ for each criterion.}
		\label{fig:sa2_smooth}
	\end{figure}

	The results of SA2 are reported in 
	Table~\ref{table:sa2_priorI} and 
	Figure~\ref{fig:sa2_smooth}. Recall that the solar 
	lighting device fails due to two independent competing 
	causes: capacitor failure (cause~1) and controller 
	failure (cause~2). The lower stress $s_1$ corresponds 
	to the temperature experienced by the device during 
	the first phase of the experiment, before the stress 
	is switched to the higher level $s_2$ at time $\tau$. 
	A lower $s_1$ means the device operates closer to its 
	normal use temperature $s_0 = 1/293$~K$^{-1}$, so 
	failures accumulate slowly and more naturally. A 
	higher $s_1$ means the device is already under 
	elevated temperature from the beginning, so failures 
	accumulate faster in the first phase.
	
	\begin{table}[H]
		\centering
		\caption{SA2: Smoothed one-variable optimal designs under Prior I 
			for nine lower stress levels $s_1 \in \{1/298, 1/303, \dots, 1/346\}$~K$^{-1}$, with $p = 0.10$, and $n = 35$. }
		\label{table:sa2_priorI}
		\renewcommand{\arraystretch}{1.2}
		\begin{tabular}{cccccc}
			\toprule
			& & \multicolumn{2}{c}{Criterion I} & 
			\multicolumn{2}{c}{Criterion II} \\
			\cmidrule(lr){3-4}\cmidrule(lr){5-6}
			$s_1$ (K$^{-1}$) & $x_1$ & $\tau^0$ & $C_1^0(\tau^0)$ & 
			$\tau^0$ & $C_2^0(\tau^0)$ \\
			\midrule
			$1/298$ & $0.1$ & $5.111$ & $0.142$ & $5.950$ & $0.059$ \\
			$1/303$ & $0.2$ & $4.720$ & $0.153$ & $5.950$ & $0.070$ \\
			$1/309$ & $0.3$ & $4.046$ & $0.170$ & $5.548$ & $0.087$ \\
			$1/314$ & $0.4$ & $3.751$ & $0.198$ & $2.923$ & $0.103$ \\
			$1/320$ & $0.5$ & $3.467$ & $0.241$ & $2.829$ & $0.121$ \\
			$1/326$ & $0.6$ & $2.935$ & $0.308$ & $2.332$ & $0.148$ \\
			$1/333$ & $0.7$ & $2.285$ & $0.404$ & $1.871$ & $0.189$ \\
			$1/339$ & $0.8$ & $2.604$ & $0.536$ & $1.433$ & $0.250$ \\
			$1/346$ & $0.9$ & $4.378$ & $0.681$ & $1.055$ & $0.321$ \\
			\bottomrule
		\end{tabular}
	\end{table}
	
	When $s_1$ is close to $s_2$, the majority of units 
	fail during the first stress phase before the stress 
	switch at $\tau$, leaving very few units to enter the 
	second phase at $s_2$. With sparse failures at the 
	higher stress level, the scale parameters 
	$\theta_1(x_2)$ and $\theta_2(x_2)$ for both causes 
	become poorly estimated. Since 
	$\log(\theta_j(x_l)) = a_j + b_j x_l$, this poor 
	estimation propagates directly into large uncertainty 
	about $a_j$ and $b_j$ for both causes, and 
	consequently inflates the posterior uncertainty about 
	$t_p(x_0)$. As a result, both $C_1^0(\tau^0)$ and 
	$C_2^0(\tau^0)$ increase monotonically with $s_1$, 
	as confirmed by Table~\ref{table:sa2_priorI}. In 
	practical terms, an experimenter who sets $s_1$ too 
	close to $s_2$ will obtain a less informative 
	experiment about the reliability of the solar lighting 
	device at use conditions, regardless of how carefully 
	the stress-change time $\tau^0$ is chosen.
	
	Regarding the optimal stress-change time, the two 
	criteria give somewhat different pictures. For 
	Criterion II, $\tau^0$ decreases monotonically from 
	$5.950$ at $x_1 = 0.1$ to $1.055$ at $x_1 = 0.9$. 
	This is physically sensible: when $s_1$ is already 
	high, most units fail quickly in the first phase and 
	there is little benefit in prolonging it. Switching 
	to $s_2$ earlier ensures that at least some units 
	survive to the second phase, providing observations 
	on $\theta_j(x_2)$ for both causes and reducing 
	posterior uncertainty about $\log(t_p(x_0))$. When 
	$s_1$ is low, on the other hand, very few failures 
	occur in the first phase and it makes sense to allow 
	more time at $s_1$ to accumulate failures from both 
	causes before switching, particularly for cause~1 
	which has a decreasing hazard rate 
	($\hat{\beta}_1 = 0.769 < 1$) and tends to produce 
	early failures.
	
	For Criterion I, a broadly similar decreasing trend 
	in $\tau^0$ is observed from $x_1 = 0.1$ to 
	$x_1 = 0.7$, however the optimal $\tau^0$ rises 
	again at $x_1 = 0.8$ and $x_1 = 0.9$. This 
	non-monotone behaviour occurs because when $s_1$ is 
	very close to $s_2$, so few units survive to the 
	second phase that the criterion surface $C_1(\tau)$ 
	becomes very flat over $\tau$ ,  there is no 
	strongly preferred stress-change time since almost 
	all failures occur in the first phase regardless of 
	when the switch happens. This flatness is clearly 
	visible in the bottom panels of 
	Figure~\ref{fig:sa2_smooth}, where the minimum of 
	$C_1(\tau)$ is shallow and poorly defined. In such 
	cases the smoothed optimum is sensitive to small 
	Monte Carlo fluctuations, which explains the 
	irregular behaviour at $x_1 = 0.8$ and $x_1 = 0.9$. 
	The criterion values at $x_1 = 0.9$ are 
	substantially larger than at lower stress levels for 
	both criteria, confirming that setting the first 
	phase temperature close to $s_2$ is informationally 
	inefficient and should be avoided in practice.

	\subsubsection{SA3: Sensitivity with Respect to Prior
		Hyperparameter Specification}
	\label{subsec:sa3}
	
\begin{figure}[tbp]
		\centering
		\includegraphics[width=0.85\linewidth, height=0.25\textheight, keepaspectratio=false]{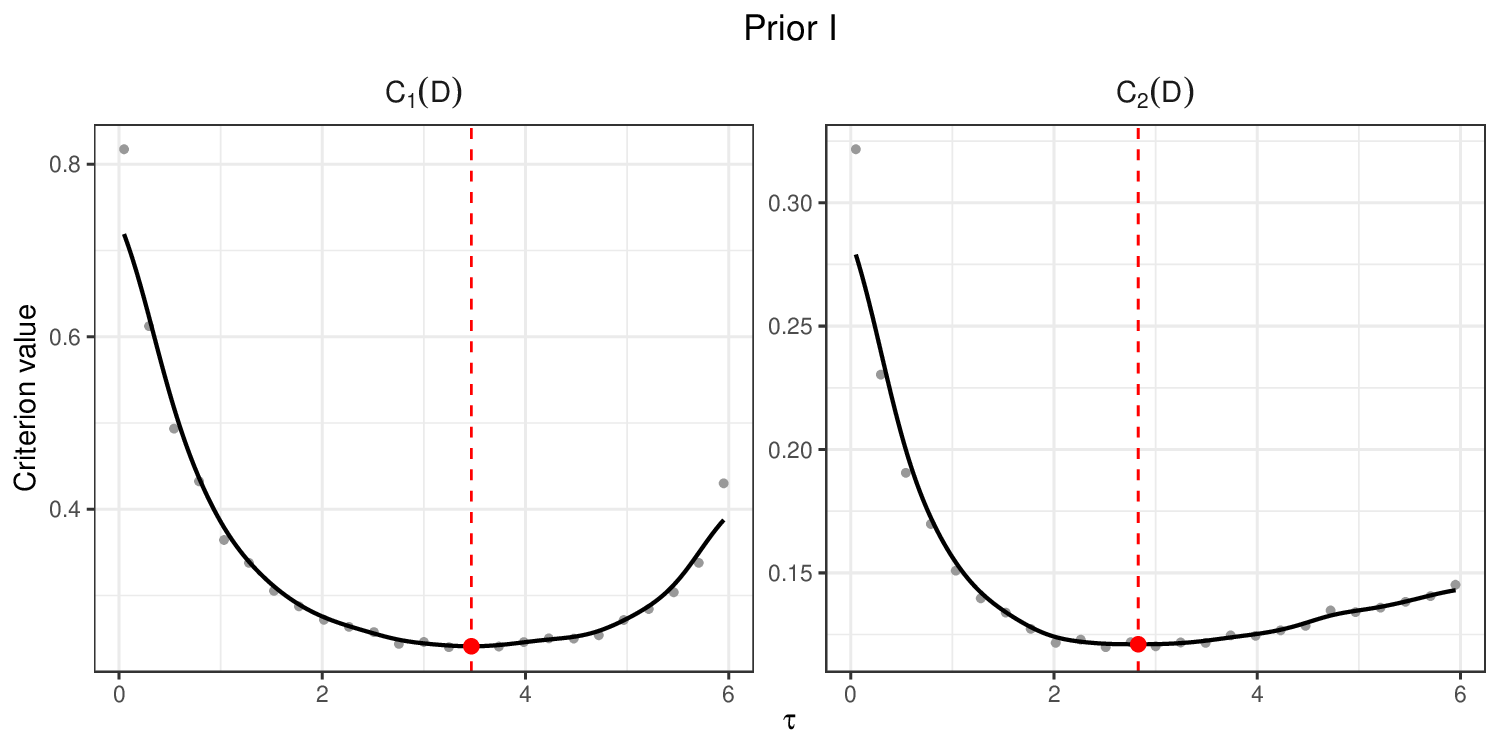}\\[4pt]
		\includegraphics[width=0.85\linewidth, height=0.25\textheight, keepaspectratio=false]{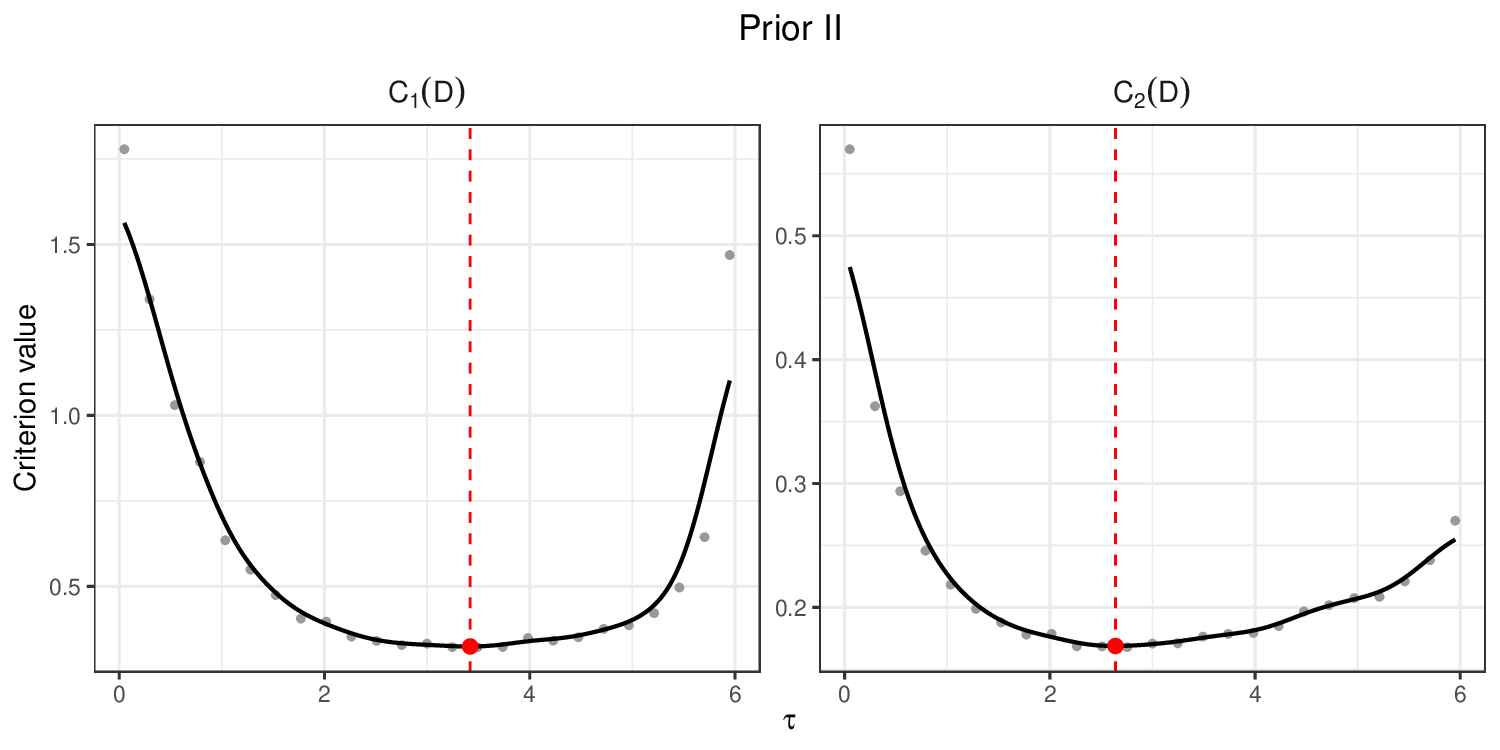}\\[4pt]
		\includegraphics[width=0.85\linewidth, height=0.25\textheight, keepaspectratio=false]{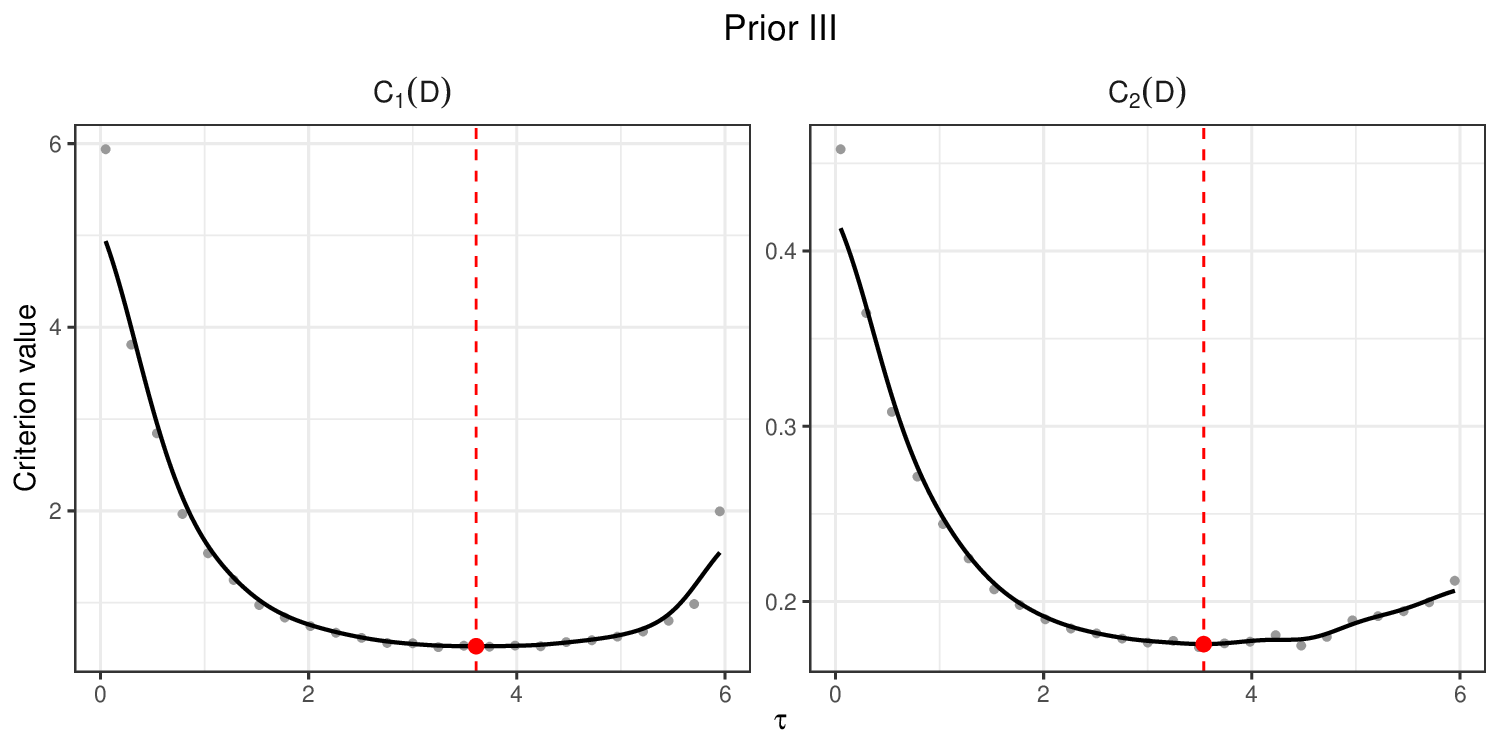}
		\caption{SA3: Smoothed criterion functions $C_1(\tau)$ and $C_2(\tau)$ under Prior choices; Prior I, Prior II and Prior III.
			Vertical dashed lines in red indicate the optimal $\tau^0$ for each criterion.}
		\label{fig:sa3}
	\end{figure}

	A primary concern in Bayesian design is whether the optimal
	design is sensitive to the choice of prior. SA3 compares
	results under Prior I, Prior II, and Prior III, which differ
	in the degree of prior informativeness and in the location
	of the prior means for the acceleration parameters, as
	defined in Section~\ref{subsec:prior}. All other parameters
	are fixed: $p = 0.10$, $s_1 = 1/320$~K$^{-1}$,
	$n = 35$. The results are reported in Table~\ref{table:sa3} and
	Figure~\ref{fig:sa3}.
	
	\begin{table}[bp]
		\centering
		\caption{SA3: Smoothed one-variable optimal designs under Prior I, Prior II 
			and Prior III, with $s_1 = 1/320$~K$^{-1}$ ($x_1 = 0.5$), $p = 0.10$, 
			$n = 35$.}
		\label{table:sa3}
		\begin{tabular}{ccccc}
			\toprule
			& \multicolumn{2}{c}{Criterion I} & \multicolumn{2}{c}{Criterion II} \\
			\cmidrule(lr){2-3} \cmidrule(lr){4-5}
			Prior & $\tau^0$ & $C_1^0(\tau^0)$ & $\tau^0$ & $C_2^0(\tau^0)$ \\
			\midrule
			Prior I & 3.467 &  0.241 &  2.829 &  0.121   \\
			Prior II & 3.077 &  0.320 & 2.527 &  0.169   \\
			%					Prior III & 3.724 &  0.517 &  3.148 &  0.175 \\
			Prior III & 3.609 &  0.526 &  3.538 &  0.176 \\
			\bottomrule
		\end{tabular}
	\end{table}
	
	As expected, the criterion values increase as we move from
	Prior I to Prior II to Prior III. Prior II is wider than
	Prior I, so greater prior uncertainty propagates into the
	preposterior variance, raising both $C_1^0(\tau^0)$ and
	$C_2^0(\tau^0)$. Prior III additionally shifts the prior
	means of the acceleration parameters upward, increasing
	the assumed activation energy and making the model believe
	that lifetime decays faster with stress. This results in
	a higher preposterior variance still, since the design must
	now accommodate greater uncertainty about both the spread
	and the location of the acceleration parameters.
	
	The behaviour of $\tau^0$ across the three priors is more
	subtle. Moving from Prior I to Prior II, $\tau^0$ decreases
	slightly for both criteria, from $3.467$ to $3.077$ under
	Criterion I and from $2.829$ to $2.527$ under Criterion II.
	When the prior is more diffuse, the experiment itself must
	be more informative to compensate for the reduced prior
	knowledge. Switching to the higher stress $s_2$ slightly
	earlier allows more failures to accumulate at the elevated
	stress level, providing more information about the
	parameters and reducing posterior
	uncertainty about $t_p(x_0)$. Moving further to
	Prior III, $\tau^0$ rises again to $3.609$ and $3.538$
	respectively. Under Prior III the model assumes stronger
	acceleration, meaning units are expected to fail more
	quickly under elevated stress. The optimal response is to
	keep units at $s_1$ longer before switching, to ensure
	sufficient failure data accumulate in the first phase before
	the stress change and to maintain the overall information from the experiment, which pushes $\tau^0$ upward. The overall range of $\tau^0$ across the three priors is modest, roughly half a unit on the time scale,
	suggesting that the optimal design is reasonably robust 
	to the prior specification considered here.
	
	\subsubsection{SA4: Sensitivity with Respect to Sample
		Size $n$}
	\label{subsec:sa4}
	
\begin{figure}[H]
		\centering
		\includegraphics[width=0.85\linewidth, height=0.25\textheight, keepaspectratio=false]{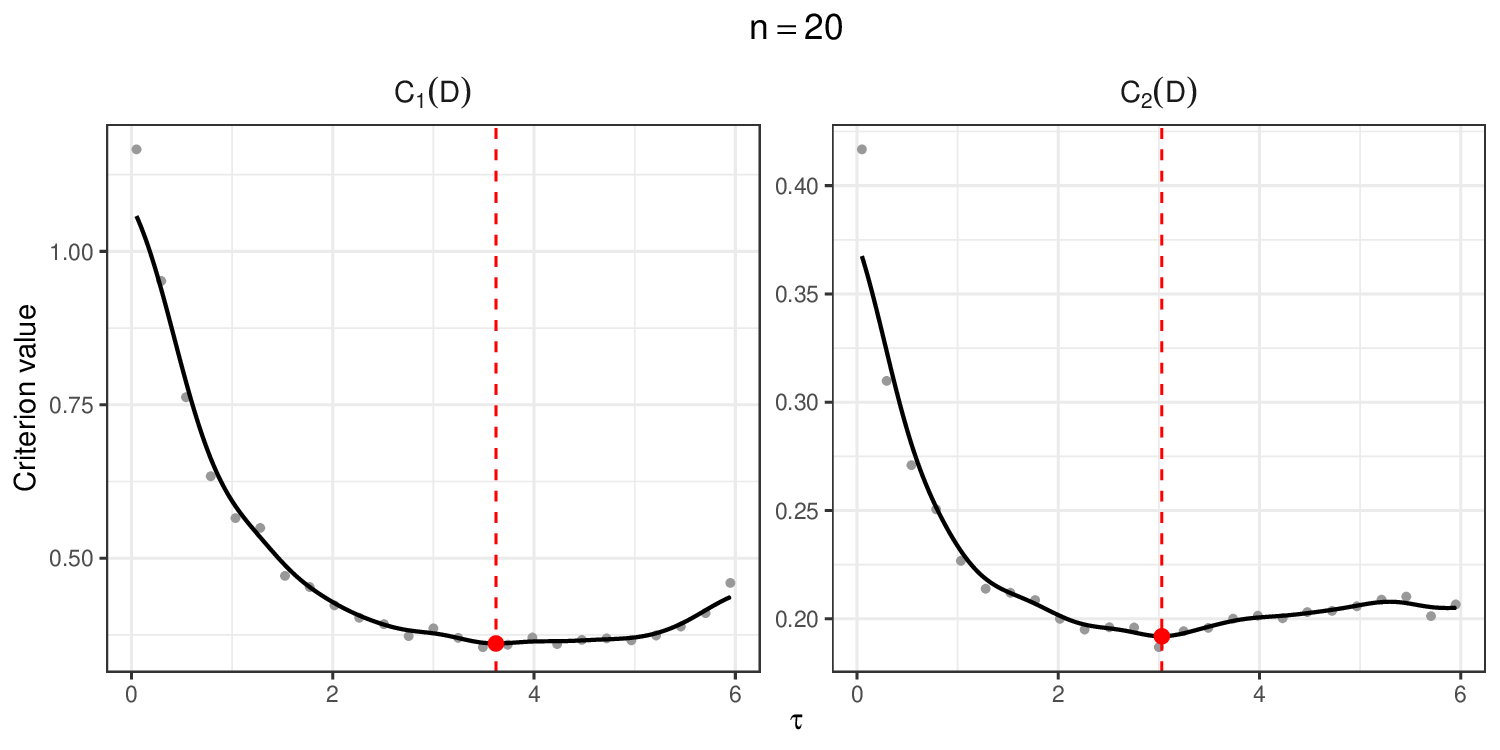}\\[4pt]
		\includegraphics[width=0.85\linewidth, height=0.25\textheight, keepaspectratio=false]{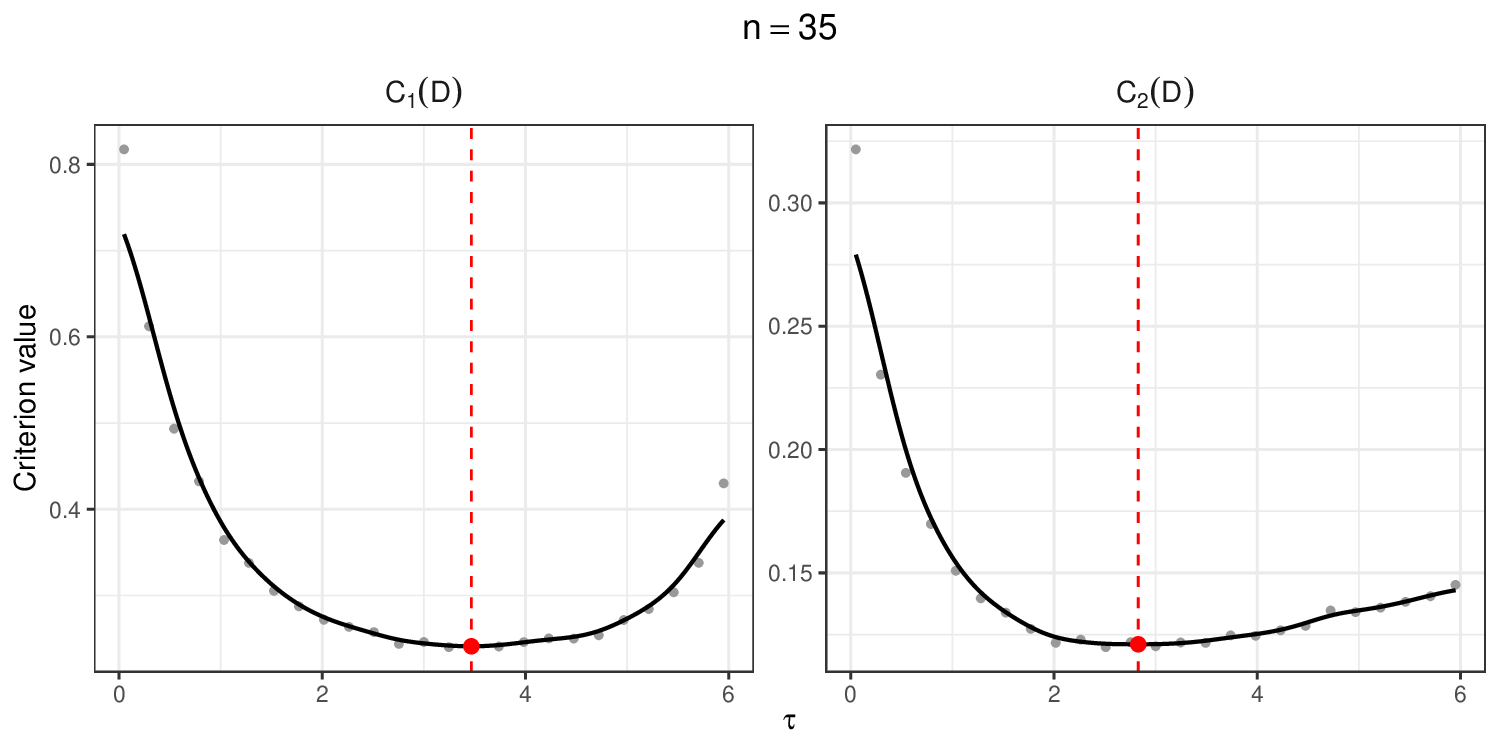}\\[4pt]
		\includegraphics[width=0.85\linewidth, height=0.25\textheight, keepaspectratio=false]{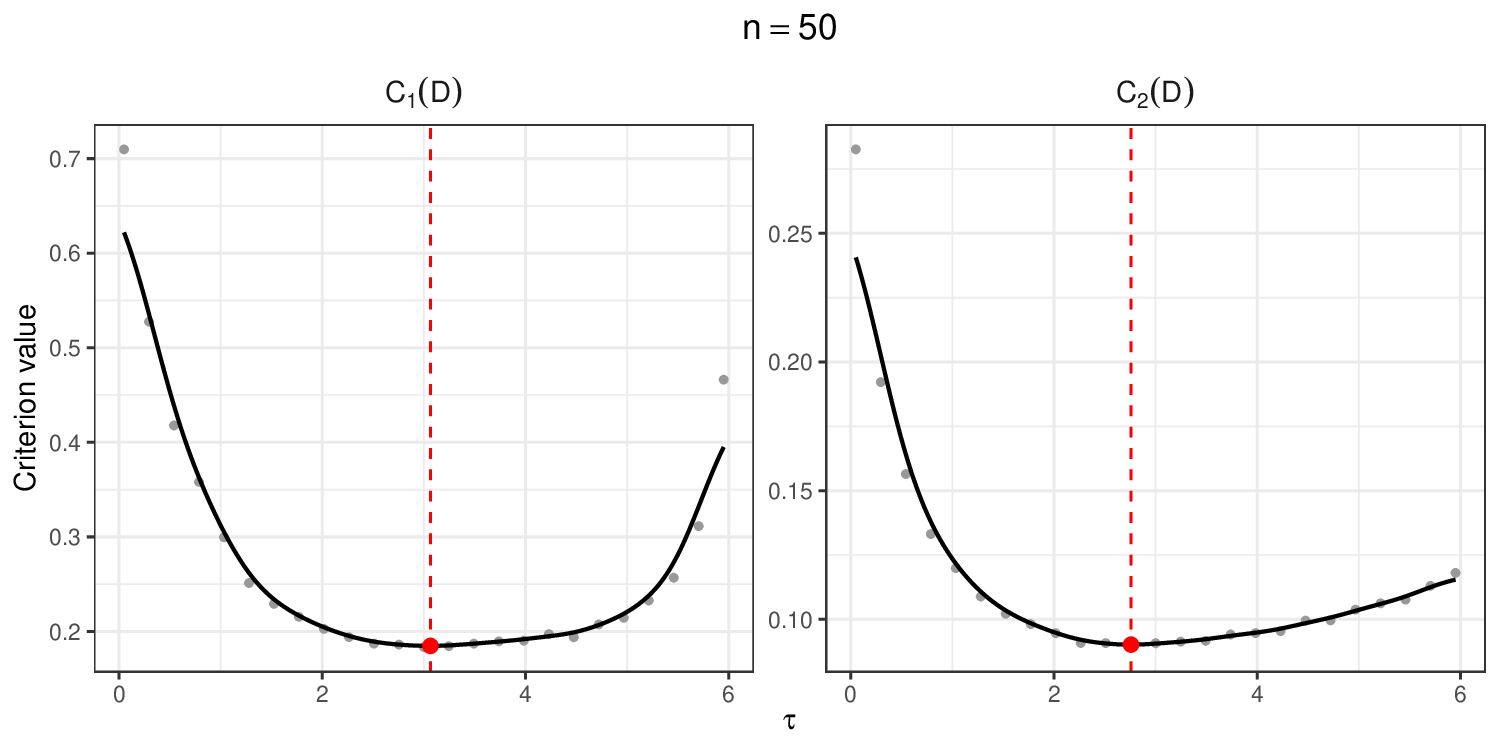}
		\caption{SA4: Smoothed criterion functions $C_1(\tau)$ and $C_2(\tau)$
			under varying sample size $n \in \{20, 35, 50\}$.
			Vertical dashed lines in red indicate the optimal $\tau^0$ for each criterion.}
		\label{fig:sa4}
	\end{figure}

	In practical reliability testing, the sample size $n$ is
	often constrained by cost, material availability, or testing
	capacity. The entire motivation for the present work is that
	the design methodology should be valid and useful precisely
	for small $n$, where asymptotic approximations are
	unreliable. SA4 examines how the optimal design and the
	achievable criterion values change as $n$ varies over
	$\{20,\; 35,\; 50\}$. All other parameters are fixed:
	$p = 0.10$, $s_1 = 1/320$~K$^{-1}$, Prior I. The results are reported in Table~\ref{table:sa4} and
	Figure~\ref{fig:sa4}.
	
	\begin{table}[bp]
		\centering
		\caption{SA4: Smoothed one-variable optimal designs under varying sample size $n \in \{20, 35, 50\}$, with $s_1 = 1/320$~K$^{-1}$ ($x_1 = 0.5$), $p = 0.10$, and Prior I.}
		\label{table:sa4}
		\begin{tabular}{ccccc}
			\toprule
			& \multicolumn{2}{c}{Criterion I} & \multicolumn{2}{c}{Criterion II} \\
			\cmidrule(lr){2-3} \cmidrule(lr){4-5}
			$n$ & $\tau^0$ & $C_1^0(\tau^0)$ & $\tau^0$ & $C_2^0(\tau^0)$ \\
			\midrule
			20 & 3.621 & 0.361 & 3.030 & 0.192 \\ 
			35 & 3.467 & 0.241 & 2.829 & 0.121 \\
			50 & 3.065 & 0.185 & 2.758 & 0.090 \\ 
			\bottomrule
		\end{tabular}
	\end{table}
	
	As $n$ increases, the posterior distribution of $t_p(x_0)$
	concentrates more tightly around the true value, and both
	$C_1^0(\tau^0)$ and $C_2^0(\tau^0)$ decrease accordingly,
	from $0.361$ and $0.192$ at $n = 20$ to $0.185$ and
	$0.090$ at $n = 50$ under the two criteria. This confirms
	that larger samples yield more precise estimates of the
	reliability characteristic of interest, as expected.
	
	The optimal stress-change time $\tau^0$ also shifts with
	$n$, decreasing from $3.621$ at $n = 20$ to $3.065$ at
	$n = 50$ under Criterion I, and from $3.030$ to $2.758$
	under Criterion II. With more units available, enough
	failures accumulate in the first phase even with a shorter
	$\tau$, so there is no need to wait as long before
	switching to the higher stress to gather information about
	the scale parameters $\theta_j(x_2)$. The earlier switch
	allows more failures to be observed at $s_2$, which is
	particularly beneficial for estimating the acceleration
	parameters $b_j$ and reducing posterior uncertainty about
	$t_p(x_0)$ at use conditions. The shift in $\tau^0$ across
	the three sample sizes is moderate, around half a unit on
	the time scale, suggesting that the design recommended for
	$n = 35$ provides a reasonable starting point even when the
	actual sample size deviates somewhat from the planned value.
	The full Bayesian computation via NUTS ensures that the
	criterion values are accurate even at $n = 20$, where
	asymptotic Fisher-information-based methods would be
	unreliable.
	
	\subsection{Two-Variable Optimal Design}
	\label{subsec:twovariable}
	
\begin{figure}[tbp]
		\centering
		\includegraphics[width=0.9\linewidth]{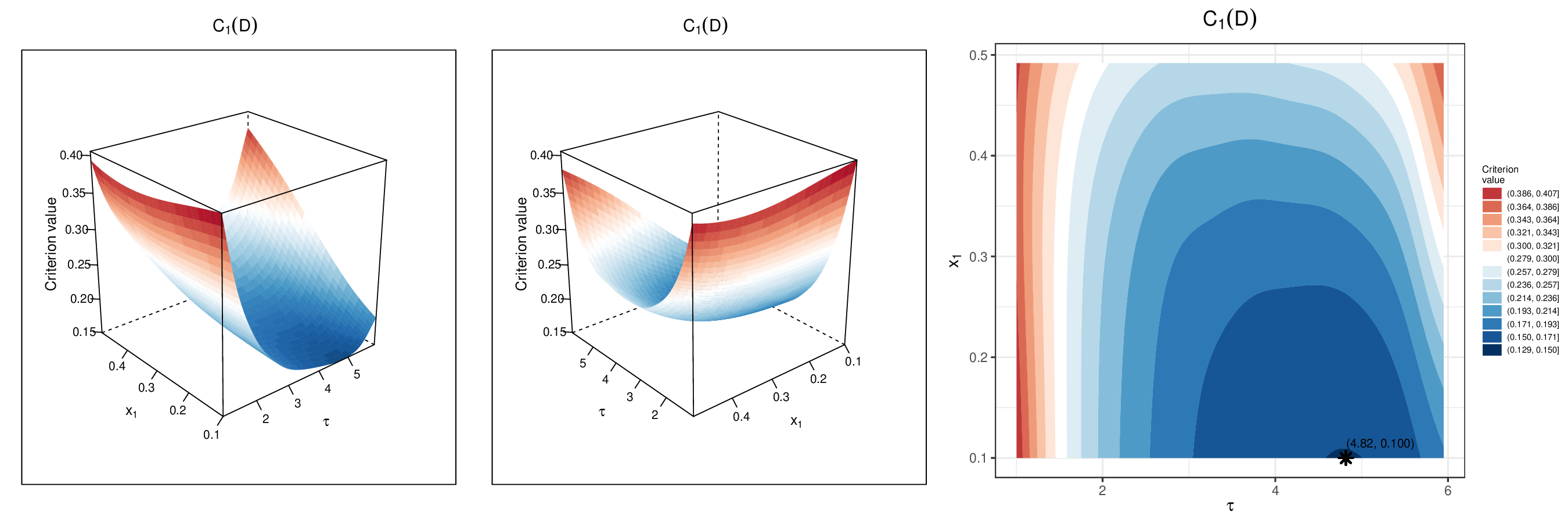}
		
		\vspace{0.3em}
		
		\includegraphics[width=0.9\linewidth]{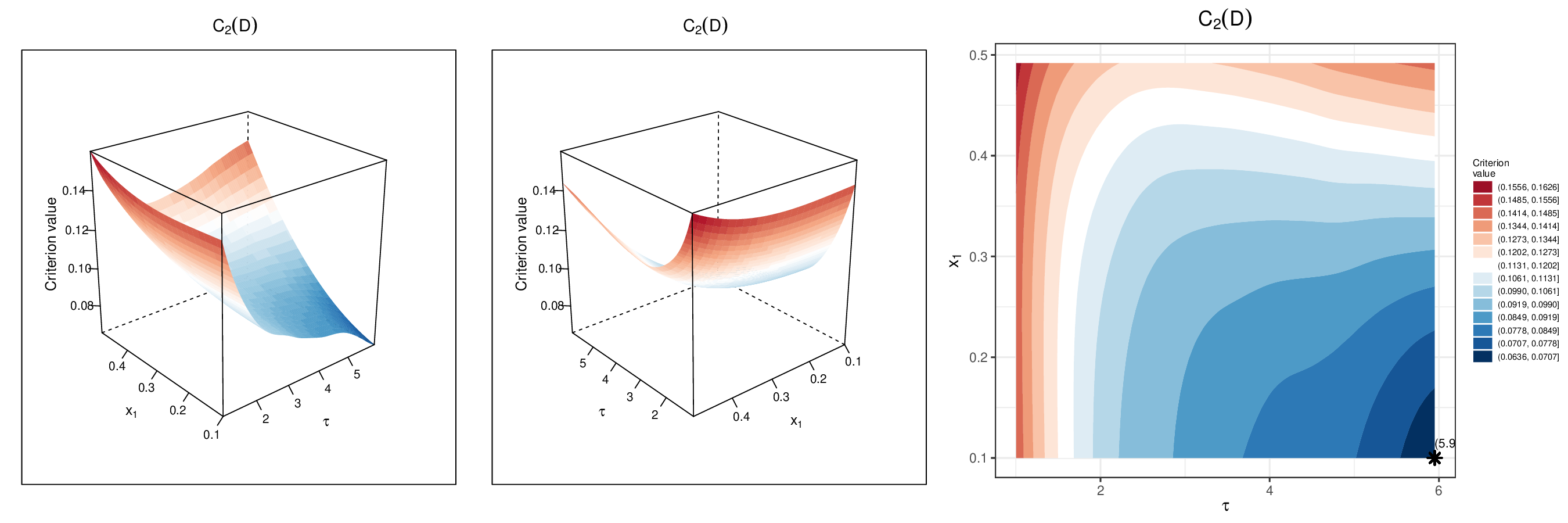}
		
		\vspace{0.3em}
		
		\begin{subfigure}{\linewidth}
			\phantom{x}
			\caption*{\textbf{Prior I}}
			\label{fig:sa2_3d_PI}
		\end{subfigure}
		
		\vspace{0.8em}
		
		\includegraphics[width=0.9\linewidth]{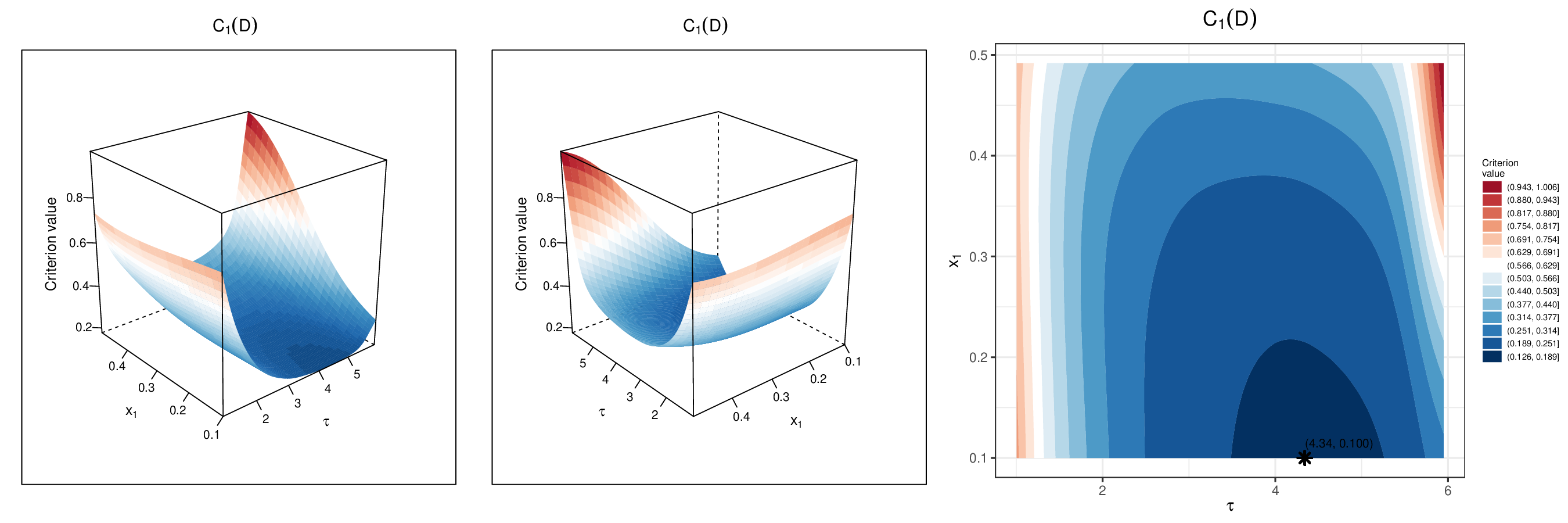}
		
		\vspace{0.3em}
		
		\includegraphics[width=0.9\linewidth]{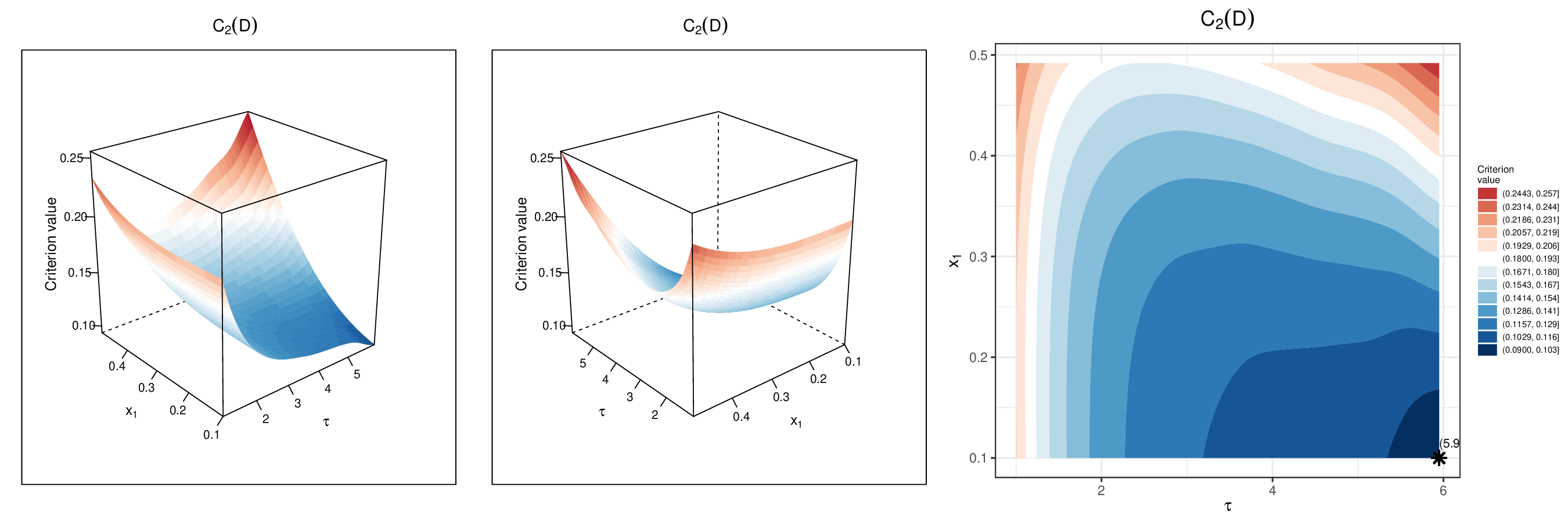}
		
		\vspace{0.3em}
		
		\begin{subfigure}{\linewidth}
			\phantom{x}
			\caption*{\textbf{Prior II}}
			\label{fig:sa2_3d_PII}
		\end{subfigure}
		
		\caption{Smoothed criterion surfaces $C_1(D)$ and $C_2(D)$ over the
			two-dimensional design space $(x_1, \tau)$ under Prior I (tighter,
			top) and Prior II (wider, bottom). Each surface is shown from two viewing angles alongside a contour plot. }%The minimum is attained at $s_1^0 = 1/298$~K$^{-1}$ ($x_1 = 0.1$) for both criteria and both priors.}}
\label{fig:sa2_3d}
\end{figure}

	The two-variable optimal design is obtained by minimising 
	$C_1(D)$ and $C_2(D)$ jointly over both $s_1$ and $\tau$ 
	using Algorithm~\ref{algo_2}, with $p = 0.10$, $n = 35$, 
	and $B = 1000$. The results are reported in 
	Table~\ref{table:twovariable_optimal} and 
	Figure~\ref{fig:sa2_3d}.
	
	\begin{table}[H]
		\centering
		\caption{Smoothed two-variable optimal designs under Prior I, Prior II, 
			and Prior III, with $p = 0.10$ and $n = 35$.}
		\label{table:twovariable_optimal}
		\renewcommand{\arraystretch}{1.2}
		\begin{tabular}{lcccccc}
			\toprule
			& \multicolumn{3}{c}{Criterion I} & 
			\multicolumn{3}{c}{Criterion II} \\
			\cmidrule(lr){2-4}\cmidrule(lr){5-7}
			Prior & $s_1^0$ (K$^{-1}$) & $\tau^0$ & $C_1^0(D^0)$ & 
			$s_1^0$ (K$^{-1}$) & $\tau^0$ & $C_2^0(D^0)$ \\
			\midrule
			Prior I & $1/298$ & $4.818$ & $0.149$ & $1/298$ & $5.950$ & $0.065$ \\
			%		Prior I  & $1/298$ & $5.111$ & $0.142$ & $1/298$ & $5.950$ & $0.059$ \\
			Prior II  & $1/298$ & $4.341$ & $0.178$ & $1/298$ &  $5.950$ & $0.093$ \\
			Prior III & $1/298$ & $4.758$ & $0.216$ & $1/298$ & $5.950$ & $0.095$\\
			\bottomrule
		\end{tabular}
	\end{table}
	
	For both criteria and all three priors, the two-variable 
	optimal design selects $s_1^0 = 1/298$~K$^{-1}$ 
	($x_1 = 0.1$), the lowest feasible stress level in the 
	experimental grid. The corresponding optimal 
	stress-change times and criterion values are given in 
	Table~\ref{table:twovariable_optimal}, and the smoothed criterion surfaces over the full design space $(x_1, \tau)$ 
	are shown in Figure~\ref{fig:sa2_3d}.
	
	This boundary optimum is consistent with the one-variable 
	findings of SA2 and has a clear physical explanation rooted 
	in the characteristics of the solar lighting device. The 
	true model parameters are estimated from data collected at 
	$s_1 = 293$~K, the normal operating temperature, at which 
	the device rarely fails under natural conditions. This 
	means the lifetime distribution at use stress has a long 
	right tail, and estimating $t_p(x_0)$ precisely requires 
	the experiment to collect failure information as close to 
	use conditions as possible. Setting $s_1$ close to 
	$s_0 = 1/293$~K$^{-1}$ achieves this: units spend the 
	first phase at a temperature near normal operation, 
	failures accumulate slowly and from both causes, and the 
	majority of units survive to enter the second phase at 
	$s_2 = 1/353$~K$^{-1}$. The second phase then provides 
	the accelerated failure data needed to identify the 
	acceleration parameters $b_j$ for both causes. This 
	two-phase balance, natural failures at low stress 
	followed by accelerated failures at high stress, is 
	what minimises posterior uncertainty about $t_p(x_0)$ for 
	this device, and it is achieved most effectively when 
	$s_1$ is as low as feasible.
	
	Setting $s_1$ higher reduces the contrast between the two 
	phases and causes more units to fail in the first phase 
	before the stress switch, leaving fewer units to enter the 
	second phase. As shown in SA2, this progressively 
	inflates both $C_1^0(\tau^0)$ and $C_2^0(\tau^0)$, 
	explaining why the criterion surface decreases 
	monotonically toward $x_1 = 0.1$ for both criteria, as 
	visible in Figure~\ref{fig:sa2_3d}.
	
	The boundary optimum at $s_1^0 = 1/298$~K$^{-1}$ is driven 
	by the strong thermal acceleration of the capacitor failure 
	mode ($\hat{b}_1 = -4.713$, Table~\ref{table:mle_par}). With 
	$\theta_1(x_l)$ decreasing sharply with stress, preserving 
	units for the high-stress phase consistently outweighs the 
	benefit of accumulating first-phase failures, and the lowest 
	feasible $s_1$ remains optimal throughout. This holds across 
	all three prior specifications. For devices with weaker 
	thermal acceleration, an interior optimum over $s_1$ would be 
	expected, requiring the full two-variable optimisation.
	
The criterion values increase from Prior~I to Prior~II to 
Prior~III, consistent with SA3, wider and more shifted 
priors propagate greater uncertainty into the preposterior 
criterion. The optimal $\tau^0$ varies moderately across 
priors under Criterion~I, from $4.341$ to $4.922$, while 
remaining at the upper boundary $5.950$ under Criterion~II 
for all three priors. The stability of $s_1^0 = 1/298$ 
across all prior specifications confirms that this finding 
is not an artefact of the prior choice but reflects the 
genuine information structure of the step-stress 
experiment for this particular device. Experimenters 
planning a similar SSALT study should therefore set $s_1$ 
as close to the use stress as operationally feasible.

\section{Conclusion}
\label{sec:conclusion}

This paper developed a Bayesian framework for planning
simple step-stress accelerated life tests when items are 
subject to two independent competing failure modes. The proposed methodology avoids 
asymptotic approximations entirely, relying instead on 
NUTS-based posterior sampling and Monte Carlo simulation 
to evaluate the preposterior variance of the $p$-th 
quantile of the lifetime distribution at use stress. 
This makes the proposed designs reliable for the 
small sample sizes that are typical in industrial 
reliability testing, as discussed in 
Section~\ref{sec:intro}, where the limitations of 
asymptotic approximations based on the Fisher 
information matrix motivated the present work.

A key contribution is the reparametrisation of the SSALT 
model parameters to $\boldsymbol{\varphi}_j = 
(t^q_{0,j}, -b_j, \beta_j)$, which admits physically 
interpretable and approximately independent prior 
elicitation from engineering knowledge. The sensitivity 
analysis confirmed that the optimal stress-change time 
$\tau^0$ is influenced by the quantile level $p$, the 
lower stress $s_1$, and the sample size $n$, but remains 
reasonably stable across the three prior specifications 
considered, with $\tau^0$ 
shifting by at most half a unit on the time scale. Designs targeting 
early-life failures favour an earlier stress switch, 
while those targeting the median lifetime favour a 
later switch, so the choice of $p$ should reflect 
the reliability characteristic of primary 
engineering concern. 

The two-variable optimal design consistently 
selects the lowest feasible lower stress level for the 
solar lighting device, a finding that is robust to all 
prior choices and reflects the genuine information 
structure of the step-stress experiment for this device.
This boundary optimum is not a limitation of the 
methodology but a substantive finding about the 
optimal use of experimental resources for this 
particular device and failure mechanism.
This is a consequence of the strong thermal acceleration 
of the capacitor failure mode ($\hat{b}_1 = -4.713$), 
for which interior optima over $s_1$ would require 
weaker acceleration behaviour. In such cases the two-variable optimisation 
framework developed here would deliver a non-trivial 
joint recommendation that the one-variable analysis 
alone could not provide.

From a practical standpoint, the results provide clear 
guidance for engineers planning SSALT experiments for 
devices of this type. Setting the lower stress as close 
to the use condition as operationally feasible maximises 
the information collected about the reliability 
characteristic of interest, regardless of which criterion 
or prior specification is used.

The proposed framework is currently developed for two 
competing risks under simple SSALT. Extension to more 
than two competing risks is feasible within the 
same reparametrisation and algorithmic structure, though 
the prior elicitation problem grows in dimension. 
Extension to multiple stress levels would be 
computationally demanding given the Monte Carlo loop over 
the design space, and is left for future work. A further extension worth pursuing is the joint 
optimisation of the stress-change time, the lower stress 
level, and the sample size under a constrained and multicriteria framework 
incorporating inspection and testing costs 
\citep{ehrgott2005multicriteria, han2015time, bhattacharyya2020multi, prajapat2026constrained}.
%\appendix
%\section{Codes used for simulation}
%\newpage
%\lstinputlisting[language=C++, caption={Stan Model}, label={lst:lst1}]{C:/Users/Kiran/Documents/.cmdstan/cmdstan-2.33.1/examples/bernoulli/bernoulli2.stan}
%\lstinputlisting[language=R, caption={R code for Posterior Estimates}, label={lst:lst2}]{C:/Users/Kiran/sciebo - Prajapat, Kiran (0U00ZN@rwth-aachen.de)@rwth-aachen.sciebo.de/RWTH/OBD_SSALT_CR/Computations/Test/Posterior_Estimates_for_n.R}

\section*{Acknowledgements}
The author thanks Prof.\ Maria Kateri (RWTH Aachen 
University) for valuable discussions during the early 
stages of this work and for hosting the postdoctoral 
position during which this research was initiated. The author acknowledges the support 
of the High-Performance Computing (HPC) systems at the RWTH Aachen University, and Newcastle 
University.

%\section*{Conflict of Interest}
%The author hereby declares that the information provided 
%here is accurate, and there are no apparent conflicts of 
%interest relevant to the content of this article.

\bibliographystyle{chicago}
\bibliography{references}

\end{document}